\documentclass[twocolumn]{aastex631}
\usepackage{hyphenat}
\usepackage{amsmath, eqnarray}
\usepackage{amssymb}
\usepackage{textcomp, gensymb}
\usepackage{soul}
\usepackage{nicematrix}
\usepackage{footnote}
\usepackage{diagbox}

\usepackage{enumitem}

\newcommand\BO{\textrm{1}}
\newcommand\BB{\textrm{2}}
\newcommand\BOa{\textrm{1a}}
\newcommand\BOb{\textrm{1b}}
\newcommand\BBa{\textrm{2a}}
\newcommand\BBb{\textrm{2b}}
\newcommand\TYC{TYC 3340-2437-1}

\newcommand{\RP}{_\textrm{RP}}

\newcommand\vsini{$v\sin i$}
\newcommand{\Angstrom}{\textrm{\AA}}
\newcommand\thestar{TYC 3340-2437-1}

\newcommand{\kms}{km\ s$^{-1}$}
\newcommand{\lijiao}[1]{\textcolor{red}{#1}}

\usepackage{hyperref}

\shortauthors{Jiao Li et al.}

\graphicspath{{./}{figures/}}

\begin{document}

\title{TYC 3340-2437-1: A Quadruple System with A Massive Star}

\author[0000-0002-2577-1990]{Jiao Li}
\affiliation{Key Laboratory of Space Astronomy and Technology, National Astronomical Observatories, Chinese Academy of Sciences, Beijing 100101, People’s Republic of
China}


\author[0000-0002-1802-6917]{Chao Liu}
\affiliation{Key Laboratory of Space Astronomy and Technology, National Astronomical Observatories, Chinese Academy of Sciences, Beijing 100101, People’s Republic of
China}

\author{Changqing Luo}
\affiliation{Key Laboratory of Space Astronomy and Technology, National Astronomical Observatories, Chinese Academy of Sciences, Beijing 100101, People’s Republic of
China}

\author[0000-0002-6434-7201]{Bo Zhang}
\affiliation{Key Laboratory of Space Astronomy and Technology, National Astronomical Observatories, Chinese Academy of Sciences, Beijing 100101, People’s Republic of
China}

\author[0000-0003-3832-8864]{Jiang-Dan Li}
\affiliation{Yunnan Observatories, Chinese Academy of Sciences (CAS), 396 Yangfangwang, Guandu District, Kunming 650216, P.R. China}
\affiliation{University of Chinese Academy of Sciences, Beijing 100049, China}

\author[0000-0002-3651-5482]{Jia-Dong Li}
\affiliation{Key Laboratory of Space Astronomy and Technology, National Astronomical Observatories, Chinese Academy of Sciences, Beijing 100101, People’s Republic of
China}
\affiliation{University of Chinese Academy of Sciences, Beijing 100049, China}

\author[0000-0001-9204-7778]{Zhan-Wen Han}
\affiliation{Yunnan Observatories, Chinese Academy of Sciences (CAS), 396 Yangfangwang, Guandu District, Kunming 650216, P.R. China}

\author[0000-0001-5284-8001]{Xue-Fei Chen}
\affiliation{Yunnan Observatories, Chinese Academy of Sciences (CAS), 396 Yangfangwang, Guandu District, Kunming 650216, P.R. China}

\author{Xiao-Bin Zhang}
\affiliation{Key Laboratory of Optical Astronomy, National Astronomical Observatories, CAS, Beijing, 100012, China}

\author[0000-0003-4511-6800]{Lu-Qian Wang}
\affiliation{Yunnan Observatories, Chinese Academy of Sciences (CAS), 396 Yangfangwang, Guandu District, Kunming 650216, P.R. China}

\author{Min Fang}
\affiliation{Purple Mountain Observatory, Chinese Academy of Sciences, Nanjing, 210023, China}

\author{Li-Feng Xing}
\affiliation{Yunnan Observatories, Chinese Academy of Sciences (CAS), 396 Yangfangwang, Guandu District, Kunming 650216, P.R. China}

\author{Xi-Liang Zhang}
\affiliation{Yunnan Observatories, Chinese Academy of Sciences (CAS), 396 Yangfangwang, Guandu District, Kunming 650216, P.R. China}

\author[0000-0002-2006-1615]{Chichuan Jin}
\affiliation{Key Laboratory of Space Astronomy and Technology, National Astronomical Observatories, Chinese Academy of Sciences, Beijing 100101, People’s Republic of
China}
\affiliation{School of Astronomy and Space Sciences, University of Chinese Academy of Sciences, 19A Yuquan Road, Beijing 100049, China}

\correspondingauthor{Jiao Li; Chao Liu}
\email{lijiao@bao.ac.cn; liuchao@nao.cas.cn}

\begin{abstract}

Hierarchical massive quadruple systems are ideal laboratories for
examining the theories of star formation, dynamical evolution, and stellar evolution. 
The successive mergers of hierarchical quadruple systems might explain the mass gap between neutron stars and black holes. Looking for light curves of O-type binaries identified by LAMOST, 
we find a (2+2) quadruple system: \thestar{}, located in the stellar bow-shock nebula (SBN). It has a probability of over 99.99\% being a quadruple system derived from the surface density of the vicinity stars. Its inner orbital periods are 3.390602(89) days and 2.4378(16) days, respectively, and the total mass is about (11.47 + 5.79) + (5.2 + 2.02) = 24.48 $M_{\odot}$. 
The line-of-sight inclinations of the inner binaries, B$_\BO$ and B$_\BB$, are 55.94 and 78.2 degrees, respectively, indicating that they are not co-planar. Based on observations spanning 34 months and the significance of the astrometric excess noise ($D>2$) in {\it Gaia} DR3 data, we guess that its outer orbital period might be a few years. If it were true, the quadruple system might form through the disk fragmentation mechanism with outer eccentric greater than zero. This eccentricity could be the cause of both the arc-like feature of the SBN and the noncoplanarity of the inner orbit.
The outer orbital period and outer eccentric could be determined with the release of future epoch astrometric data of {\it Gaia}.

\end{abstract}

\keywords{Close binary stars (254), Massive stars (732), Spectroscopic binary stars (1557), Eclipsing binary stars (444), Interstellar clouds (834)}

\section{Introduction} \label{sec:intro}
Massive stars are progenitors of black holes (BHs) and neutron stars (NSs). Despite more and more BHs and NSs being found by the observations of LIGO, Virgo, X-ray telescopes, and radio telescopes, there still is a mass gap between NSs and BHs in the range of about $2-5$ ${M_{\odot}}$ (e.g. \citealt{Farr2011ApJ...741..103F}). It is ambiguous whether the gap is caused by observation bias or as a result of massive stellar evolution \citep{Abbott2019ApJ, Fryer2012ApJ}. In general, the evolution of a massive OB star with a mass of 8-25~$M_\odot$ is thought to possibly experience a hydrogen-rich type IIP supernovae and leave behind a neutron star \citep{Smartt2009ARA&A, Vink2021}. However, most of the massive stars are in binaries, and over 21\% of the massive stars probably lead to a merger, which could deeply influence their evolution route and fate \citep{Sana2012Sci}. A massive wide binary with a period longer than 10 years would also produce a triple-ring structure similar to that of SN 1987A \citep{Morris2007Sci}. 

If massive stars were members of quadruple systems, their evolution would be more exotic. The successive mergers of hierarchical quadruple systems might explain the mass gap between NSs and BHs \citep{Hamers2021MNRAS.506.5345H}.
In the 3+1 quadruple system, the mass gap probably arises if two NSs in the innermost orbit merge due to secular evolution forming a mass gap object, and becoming a triple system. The resulting inner binary may successively merge due to continued secular evolution, leading to the formation of a BH outside the mass gap~\citep{Safarzadeh2020ApJ...888L...3S}. In the 2+2 quadruple system, the merger remnant of an inner binary probably receives a recoil kick from the anisotropic emission of gravity waves, triggering a dynamical interaction where another merge would happen (for details, see \citealt{Fragione2020ApJ...895L..15F, Hamers2021MNRAS.506.5345H}). They are ideal laboratories to study stellar formation and evolution in a dynamically complex environment where the constituent binary stars can be close enough to interact with each other. It can also be easy to imagine that, at a specific condition for 2+2 quadruple, the less massive component binary might achieve a certain velocity to be a runaway binary if the more massive component merged and exploded as a supernova \citep{Gaoyan2019MNRAS}.

Currently, there are three main categories of formation theories for multiple systems: turbulent (core and filament) fragmentation, disk fragmentation, or dynamic interaction. The turbulent fragmentation model presumes that multiple systems generate from over-densities that develop and collapse within molecular filaments, producing initially widely separated systems ($> 500$ au). The disk fragmentation model posits that multiple systems arise from a massive accretion disk, forming initially smaller separations ($< 500$ au), and similar orientations. The dynamic interaction mode can establish gravitational bonds among nearly independently single stars at large distances through energy and angular momentum exchange with the surrounding gas cloud or individual circumstellar disks. (\citealt{Offner2022arXiv} and references therein). So, the semi-major axis of the outer orbit in quadruple systems and the orientations of their inner binaries could serve as possible indicators for discerning the formation mechanisms of multiple systems, particularly in distinguishing between the turbulent and disk fragmentation models~\citep{Lorenzo2017A&A}. Moreover, the dynamic interaction mode possesses the capability to rearrange the hierarchy and multiplicity of systems formed via turbulent or disk fragmentation.

Despite the fraction of quadruple system being about 38\% among O-type stars \citep{Moe2017ApJS}, only a handful of massive quadruples are resolved with orbital parameters (e.g. \citealt{Lorenzo2017A&A, Southworth2022, Naze2022MNRAS}). Detection and characterization of a quadruple can be challenging. It is very difficult to detect eclipse light curves, ellipsoidal light curves, or radial velocity variations for both inner binaries of a quadruple, especially, for the quadruple that inner binaries have distinct orbital inclinations.

We discover a quadruple system by cross-matching O-type stars found by the Large Sky Area Multi-Object Spectroscopic Telescope survey (LAMOST: \citealt{Liuzhicun2019ApJS, Liguangwei2021ApJS}) with the light curve of Transiting Exoplanet Survey Satellite (TESS: \citealt{Ricker2015JATIS}): \TYC{}. 
The system has two inner binaries. 
One is composed of O+B type stars, and the other is composed of B+A type stars. 
Its distance is about 2.31 kpc (see Section~\ref{sec:sed_distance}). 
The variation of the radial velocity of LAMOST Medium Resolution Survey ( LAMOST-MRS: \citealt{Liuchao2020a, Zhangbo2021ApJS}) is larger than 200 ${\rm kms^{-1}}$. 
\thestar{} has been identified as G151.9227-00.5651, located within the stellar bow-shock nebula (SBN), by \cite{Kobulnicky2016}. SBN is an arcuate structure created by the interaction between the stellar wind and the surrounding interstellar medium (ISM). \TYC{} would be an exotic object to study the evolution of massive stars.

\section{TESS light curve} \label{sec:tesslc}

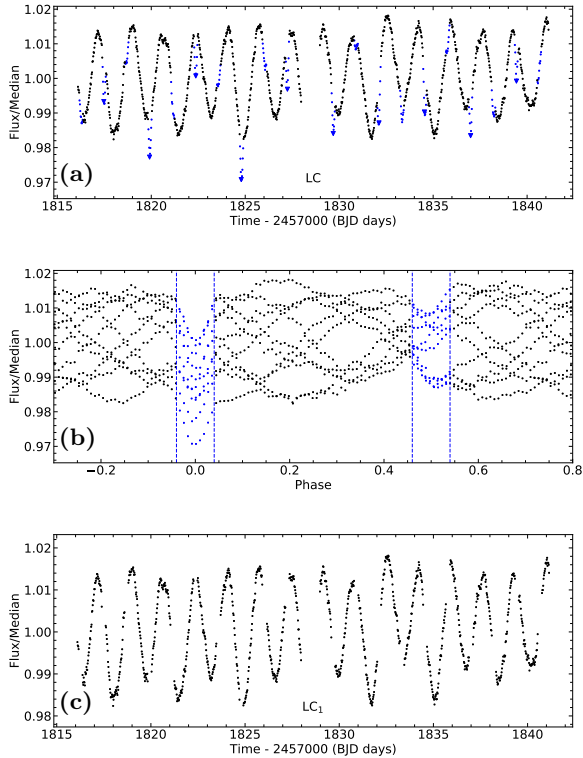
\begin{figure}[htpb]
    \centering
    \begin{picture}(200,300)
        \put(-15,200){\includegraphics[width=0.45\textwidth]{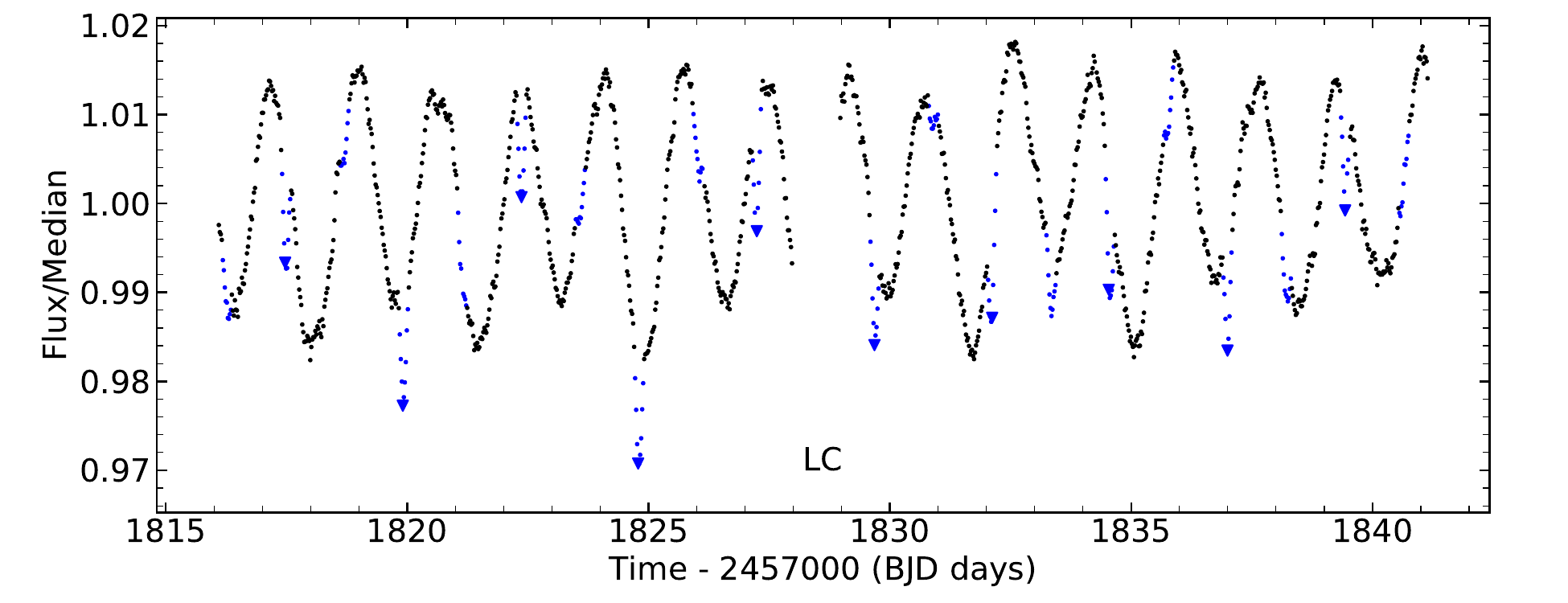}}
        \put(10,220){\textbf{(a)}}
        \put(-15,100){\includegraphics[width=0.45\textwidth]{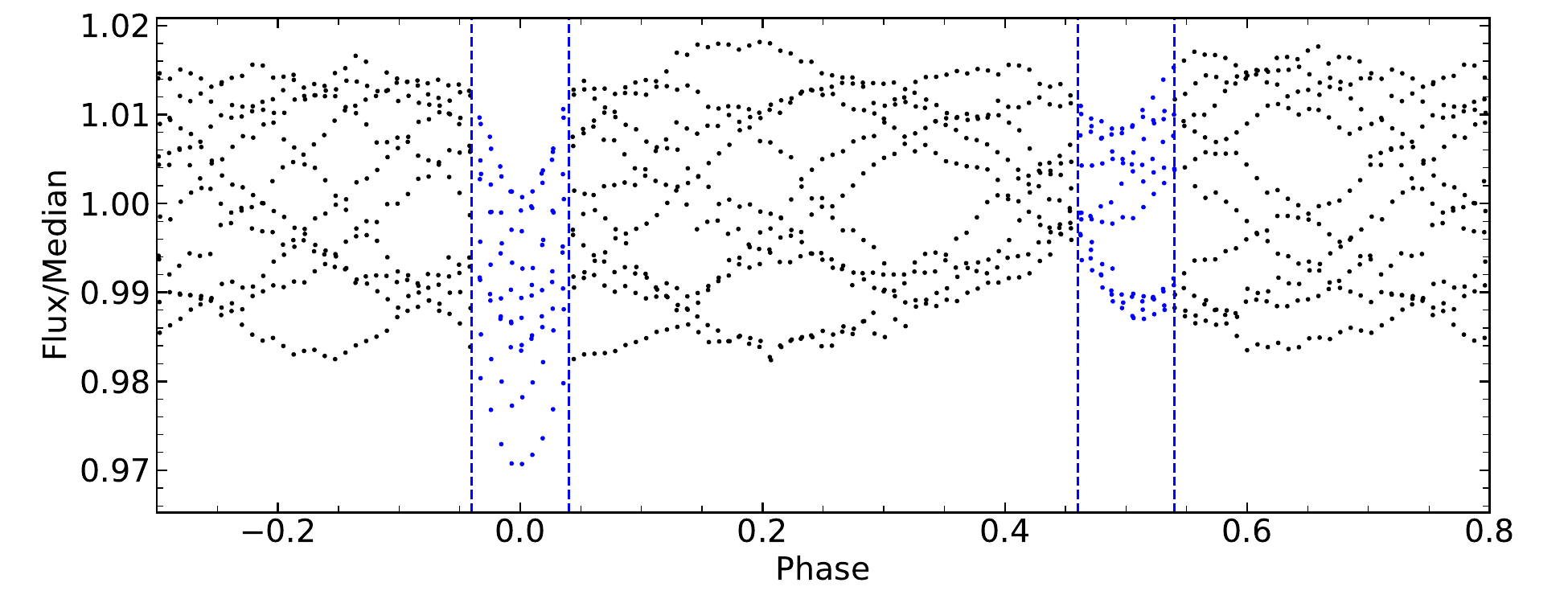}}
        \put(10,120){\textbf{(b)}}
        \put(-15,0){\includegraphics[width=0.45\textwidth]{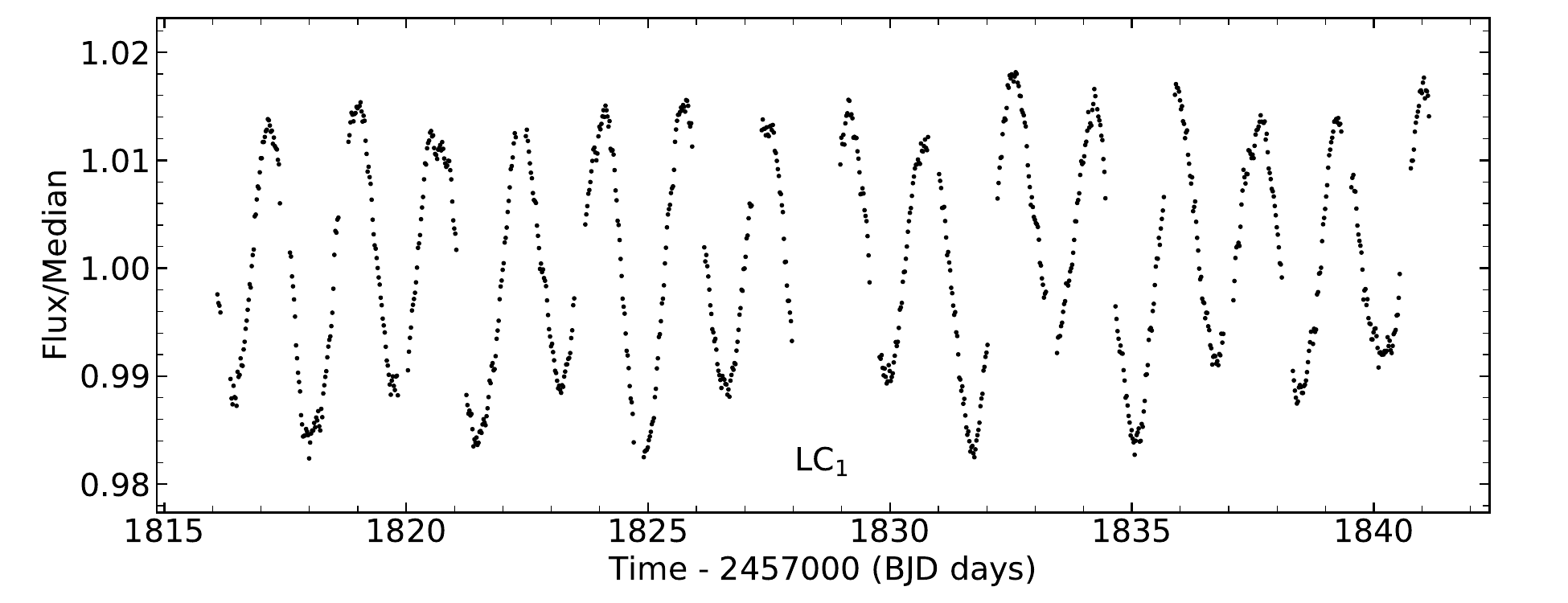}}
        \put(10,20){\textbf{(c)}}
    \end{picture}
    \caption{TESS light curve. {\it Panel} (a) shows the LC of SAP\_FLUX extracted by the MIT Quick Look Pipeline \citep{Huang2020RNAAS}; the blue triangles are the minimum points of the narrow ``valley"; the blue dots are the points in the phase of ($-0.04 < \phi < 0.04$) or ($0.46 < \phi < 0.54$) when the LC is folded with a period of 2.43781 days. {\it Panel} (b), the phase is folded with a period of 2.43781 days and to the transit time of 2458817.49056 BJD. {\it Panel} (c), the LC$_{\rm 1}$ shows the removal of the blue points in the {\it Panel} (a) 
    }
    \label{fig:tesslc_ini}
\end{figure}

\thestar{} was observed by TESS in Sector 19 with a cadence of around 30 minutes. The light curve (hereafter, we denote LC as the TESS light curve of the object) has been extracted by using the MIT Quick Look Pipeline \citep{Huang2020RNAAS}, as shown in panel (a) of Figure~\ref{fig:tesslc_ini}.

A Lomb-Scargle periodogram of LC shows a domination peak 
at $1.69759 (59)$ day (detail see Section~\ref{sec:periodogram}). 
It is similar to an ellipsoidal light curve by folding LC into twice this peak period of 3.39518 days. In addition, we have observed ten narrow valleys in LC, which are represented as blue dots with blue triangles in Figure~\ref{fig:tesslc_ini} (a). These valleys appear to have arisen due to the eclipsing of a binary system.
We fit each of the narrow valleys using the Gaussian function and obtain the conjunction points of the eclipsing binary, shown as the blue triangles of Figure~\ref{fig:tesslc_ini} (a). Then, fitting the times of the conjunction points with a linear function,
\begin{equation}
    t_i = P_2 \times i + T_{\rm c2}
\end{equation}
where $i = [0, 1, 2...10]$, $t_i$ is the $i$th conjunction time.
We obtain a period $P_{\rm 2} = 2.43781 (16)$ days and a transit time 
$T_{\rm c2} =$ 2458817.49056(75)
BJD for the eclipsing binary.

\begin{figure*}[htpb]
   \centering
    \begin{picture}(250,400)
    \put(-100,210){\includegraphics[scale=0.5, angle=0]{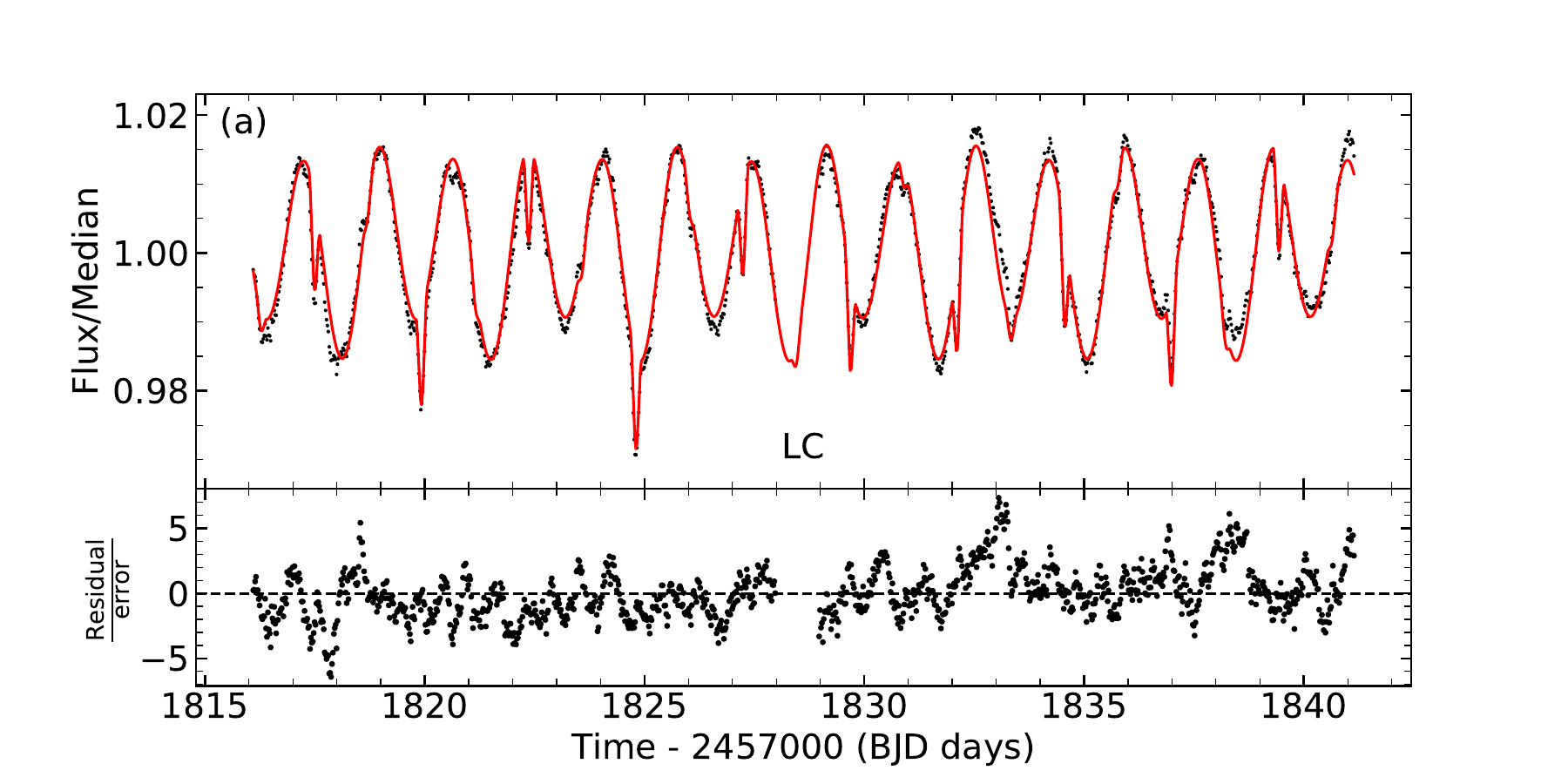}}
    \put(-100,-10){\includegraphics[scale=0.35, angle=0]{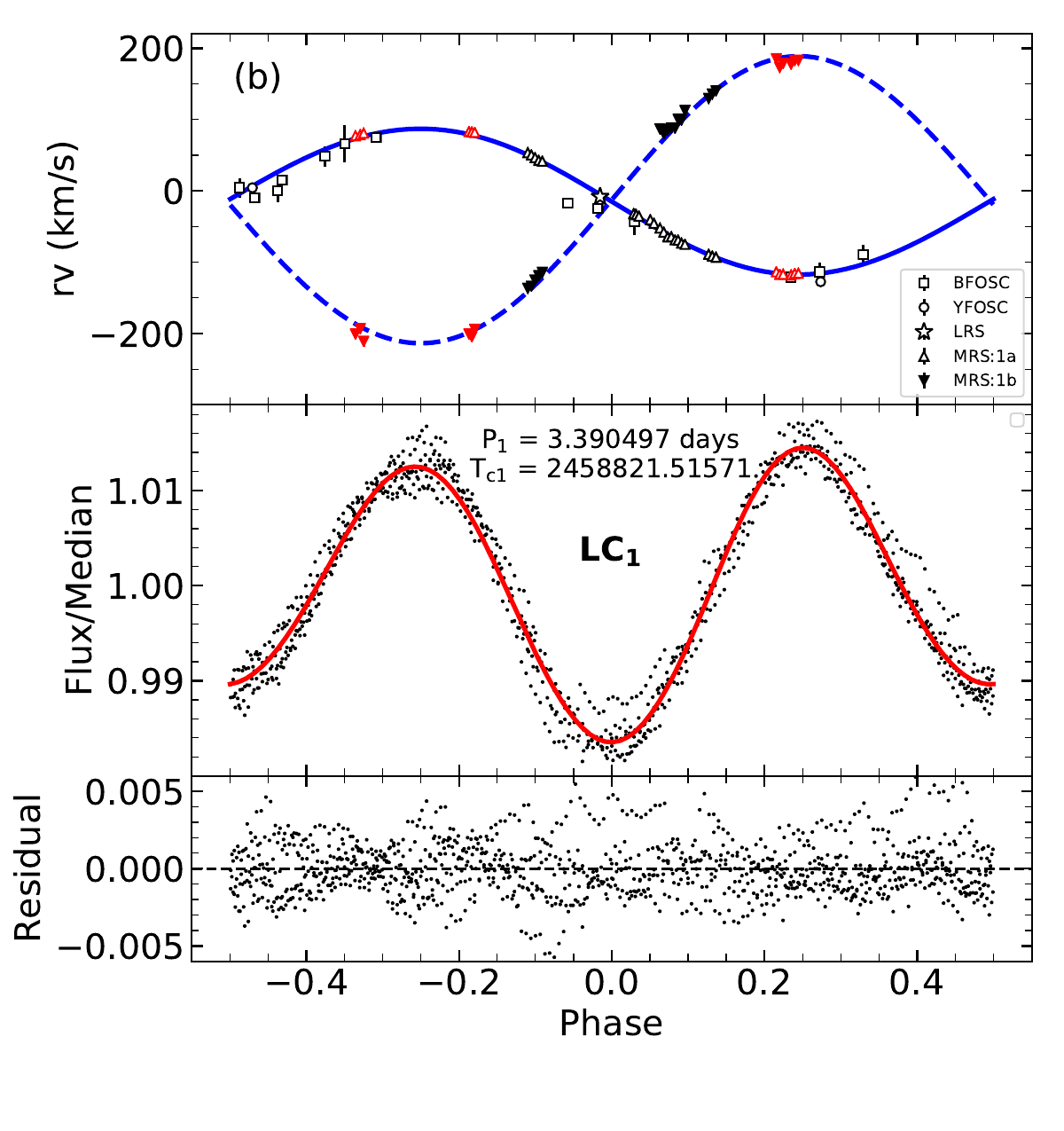}}
    \put(130,-10){\includegraphics[scale=0.35, angle=0]{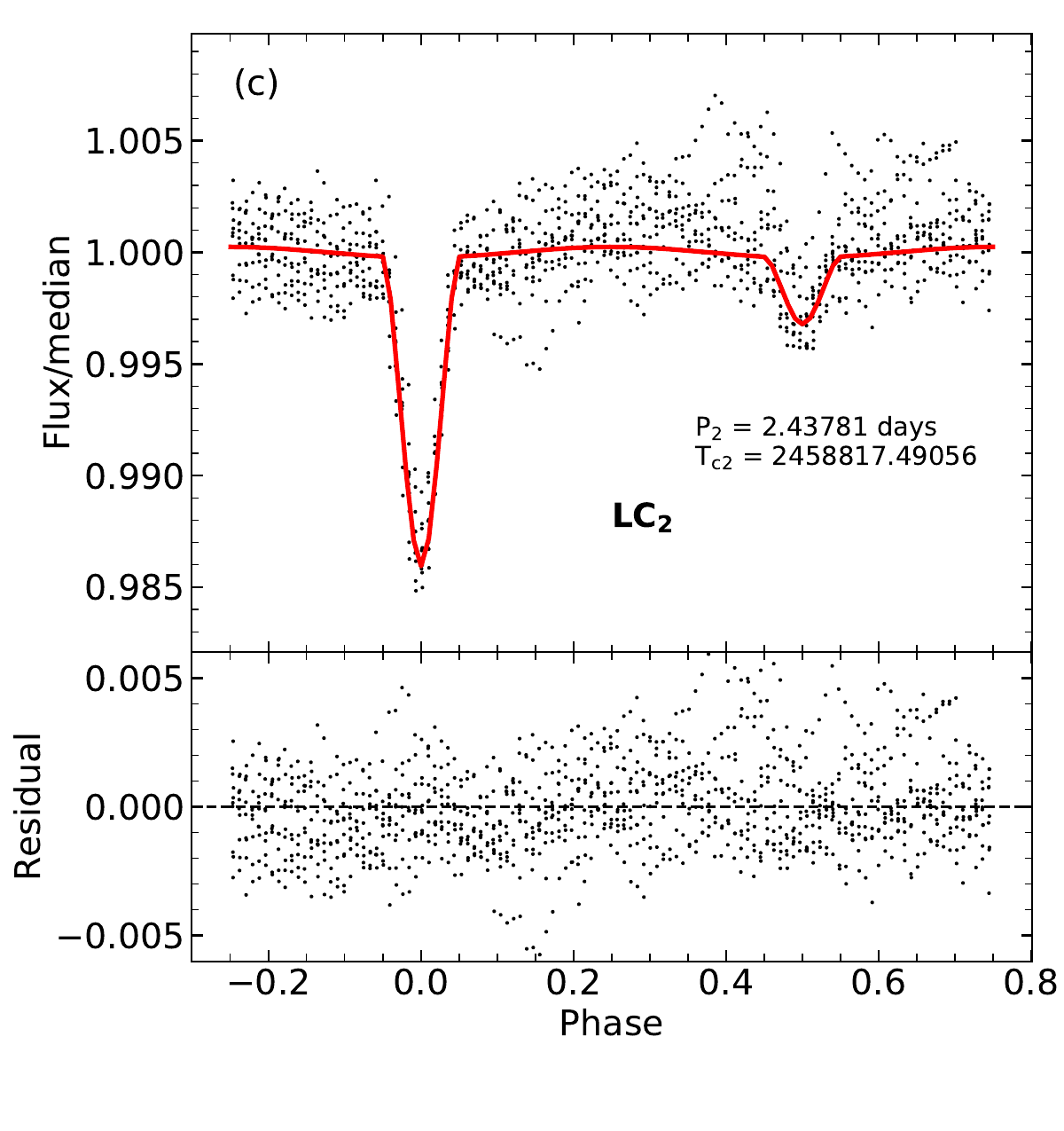}}
    \end{picture}
\caption{
\textit{Panel} (a) shows the TESS light curve (LC, dark points) of SAP\_FLUX extracted by using the MIT Quik Look Pipeline \citep{Huang2020RNAAS}; the red lines are light curve calculated by \texttt{ellc} \citep{Maxted2016}.
\textit{Panel} (b) shows that the ellipsoidal light curve of binary B$_\BO$ (LC$_\BO$) and corresponding radial velocity curve, both of which are folded by a period of 3.390497 days. In the top panel, blue solid and dash lines are Kepler orbital radial velocities of binary B$_\BOa$ and binary B$_\BOb$, respectively; the markers of the open square, dot, and star are radial velocities measured from spectra of BFOSC, YFOSC, and LAMOST-LRS ($R \sim 1800$), respectively. The open and solid triangles stand for radial velocities of star B$_\BOb$ to star B$_\BOa$, which are measured from spectra of LAMOST-MRS ($R\sim 7500$). The red triangles stand for the phases where the spectra are used to measure the radius ratio of star B$_\BOb$ to star B$_\BOa$. In the middle panel, the gray dots donate the light curve removed the data near the eclipses of binary B$_{\BB}$. The bottom panel shows the folded residual of the TESS LC. \textit{Panel} (c) shows the relative eclipse light curve of binary B$_{\BB}$ (LC$_\BB$) and folded by a period of 2.43781 days, which is obtained by dividing the TESS LC into the model light curve of binary B$_\BO$. The remarkable deviations at the phase of $\sim 0.4$ of LC$_\BB$ might be attributed to the de-trending process of the TESS LC, which can be seen at the edges of \textit{Panel} (a). \lijiao{\textit{Note}: the residuals of \textit{Panel} (b) and (c) are the same residual of \textit{Panel} (a) but folded with different periods. }
}\label{fig:tesslc}
\end{figure*}

After folding LC with $P = 2.43781$ days and $T_{\rm c} = 2458817.49056$ BJD (see Figure~\ref{fig:tesslc_ini} b), and excluding the data (the blue dots in Figure~\ref{fig:tesslc_ini} b) in phases of $(-0.04<\phi < 0.04)$ or $(0.46<\phi < 0.54)$, a new light curve (LC$_{\rm 1}$) without eclipses is obtained (see Figure~\ref{fig:tesslc_ini} c). LC$_{\rm 1}$ is folded into a period of $P_\BO=3.390497$ days, which is derived by fitting the radial velocity curve (see Section~\ref{sec:rv1}). The resulting light curve exhibits a clear sine-like feature, which is indicative of the ellipsoidal effect (as shown in the middle panel of Figure~\ref{fig:tesslc} b). By dividing LC by a model ellipsoidal light curve (or a smoothed light curve of LC$_{\rm 1}$), the eclipsing light curve (LC$_{\rm 2}$) can be obtained (see Figure~\ref{fig:tesslc} c). 

\begin{figure}[hb]
    \centering
    \includegraphics[width=0.45\textwidth]{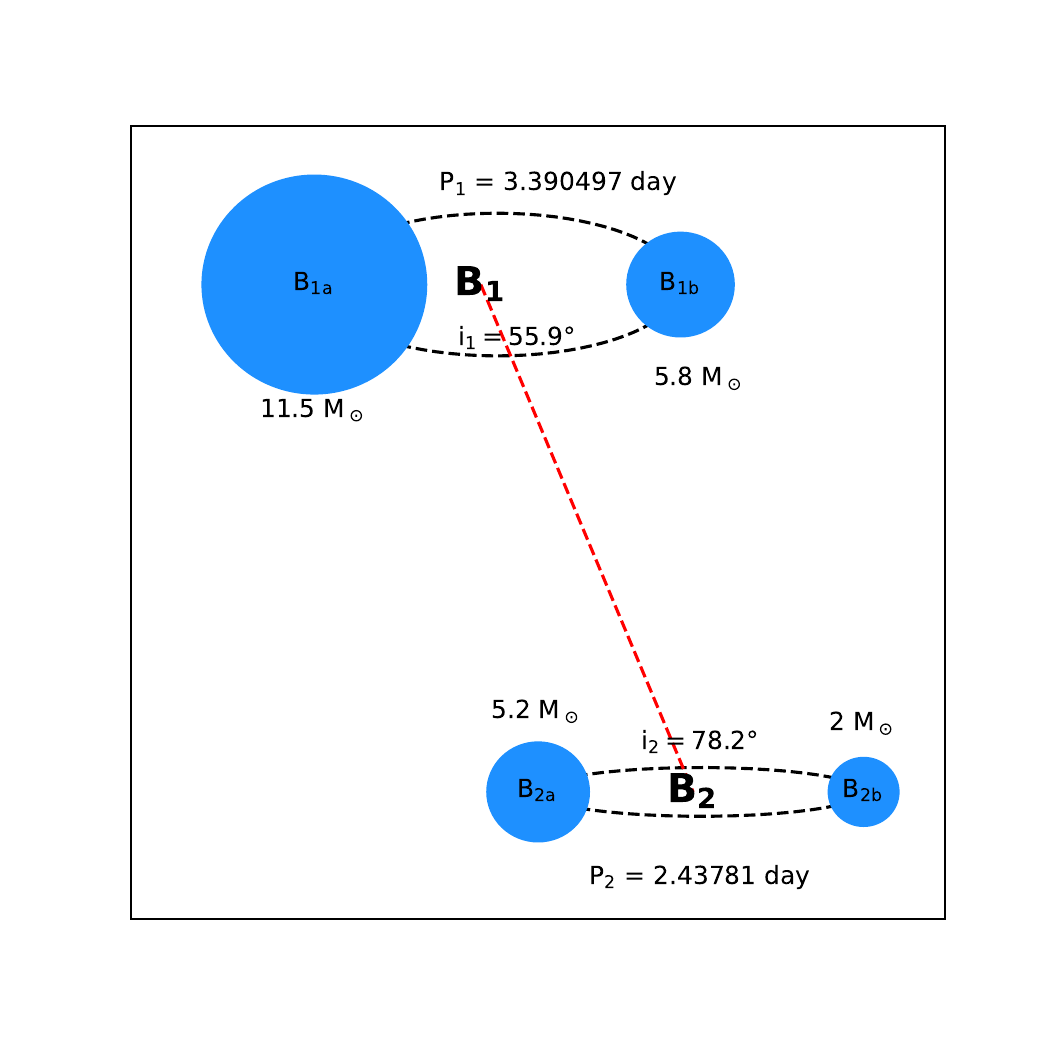}
    \caption{The schematic configuration of \thestar{}.}
    \label{fig:schematic}
\end{figure}

It is highly unlikely that the periods of LC$_\BO$ and LC$_\BB$ originate from a triple system since building a stable triple system with these two periods is extremely difficult \citep{Georgakarakos2008CeMDA, Harrington1975AJ.....80.1081H}. 
Instead, they may be contributed to a binary system in which rotation and binary orbit are asynchronous. In this case, the periods of the ellipsoidal effect and radial velocities should be 3.390497 and 2.43781 days, respectively. Alternatively, they may have arisen from two isolated binary systems or a 2+2 quadruple system.
This issue can be figured out by the radial velocity curve of the object.
In the following, we will see that LC$_{\rm 1}$ is generated from the ellipsoidal effect of an inner binary with O+B type stars (referred to as binary B$_\BO$ hereafter) of a  quadruple system, while LC$_{\rm 2}$ is induced by the other inner binary consisting of B+A type stars (referred to as binary B$_\BB$ hereafter). We define their orbital periods as $P_\BO$ and $P_\BB$. The schematic configuration of \thestar{} is shown in Figure~\ref{fig:schematic}.

\section{spectra and radial velocity curve}\label{sec:rv}

The spectra of \thestar{} have been observed through LAMOST low-resolution survey (LAMOST-LRS) and the binarity and exotic star project of LAMOST-MRS \citep{Lijiao2023ChPhB..32a9501L}. LAMOST is a reflecting Schmidt telescope with an effective aperture of about 4 m, located at the Xinglong Station of the National Astronomical Observatories of the Chinese Academy of Sciences \citep{Cuixiangqun2012RAA, Zhaogang2012RAA}. However, the radial velocities measured by the spectra were inadequate to derive a reliable period when we found the star interesting, and could not cover all phases of the orbital period $P_\BO$ (as shown by the star and triangle markers in the top panel of Figure~\ref{fig:tesslc} b ). To address this issue, follow-up spectra were observed by using the BFOSC E9+G10 instrument of the Xinglong 2.16-m telescope \citep{Zhaoyong2018RAA} and the YFOSC E9+G10 instrument of the Lijiang 2.4-m telescope (observation and data reduction see Section~\ref{sec:foscspec}).

\subsection{Radial Velocity of Star B$_{1a}$}\label{sec:rv1}

The radial velocities  (see Table~\ref{tab:rv}) are measured using cross-correlation function (CCF) with OSTAR2002 grid of TLUSTY \citep{Lanz2003ApJS}, implemented through the package of \texttt{laspec} \citep{Zhangbo2020ApJS, Zhangbo2021ApJS}. We subsequently employed a Markov Chain Monte Carlo (MCMC) approach to determine the orbital parameters of the binary with these radial velocities. The posterior distribution is 
\begin{equation}\label{eq:proprv}
   \begin{split}
    \ln[\mathcal{P}(\theta_{\BO v} \mid v_\BOa)] &\propto \ln [\mathcal{P}(v_\BOa \mid \theta_{\BO v})] \\
    &= -\frac{1}{2}\sum_{n=1}\frac{[v_{n}-v(t_n; \theta_v)]^2}{\sigma^2_{v_n}+s^2}\\ &-\frac{1}{2}\sum_{n=1}\ln[2\pi(\sigma_{v_n}^2+s^2)]
  \end{split}
\end{equation}
where $v_{n}$ and $\sigma_{v_n}$ are the radial velocities and errors; $\theta_{\BO v} = (P_\BO, T_{\rm c\BO}, K_{\BOa}, v_{\gamma\BO}, \sqrt{e_\BO}\cos\omega_\BOa, \sqrt{e_\BO}\sin\omega_\BOa)$;  $K$, $e$, $\omega$, and $v_{\gamma}$ are radial velocity semi-amplitude, eccentricity, the longitude of periastron, and systematic radial velocity; $v(t_n; \theta_v)$ is the Kepler orbital radial velocity (calculated by a python package \texttt{radvel} of \citealt{Fulton2018PASP..130d4504F}) of star B$_{\BOa}$ at a given time $t_n$; $s$ is an additional ``jitter” variance that may absorb all kinds of modeling errors such as errors of the instrument, wavelength calibration, and even radial velocity model.

\begin{table}
    \centering
    \begin{tabular}{c|c}
    \hline\hline
         $P_\BO$ (day) & 3.390497(30)\\
         $T_{\rm c\BO}$ (BJD-2457000) & $1821.503\pm0.010$\\
         $K_{\BOa}$ (\kms) & $105.4\pm1.3$\\
         $\sqrt{e_\BO}\cos\omega_\BOa$ & $0.149\pm0.046$ \\
         $\sqrt{e_\BO}\sin\omega_\BOa$ & $-0.091\pm0.052$\\
         $ v_{\gamma\BO}$ (\kms) & $-16.7\pm1.2$\\
        ``jitter” & $3.55\pm0.79$\\
    \hline
    \end{tabular}
    \caption{The orbital parameters of binary B$_\BOa$ fitted solely by radial velocity curve of star B$_\BOa$.}
    \label{tab:fit_rv1}
\end{table}

Our analysis yielded a period of 3.390497(30) days, which is consistent with the period (3.39518 days) derived from the periodogram of the light curve (see Section~\ref{sec:tesslc}). The period of 3.390497(30) is adopted as $P_\BO$ because it is more precise, thanks to the radial velocities collected over a longer observation time. The other parameters can be found in Table~\ref{tab:fit_rv1}. Upon folding the radial velocity curve and LC$_\BO$ with the period of 3.390497 days, we find that the folded LC$_\BO$ exhibits two peaks, with the peak near phase 0.25 being slightly higher than the one near phase -0.25 (see Figure~\ref{fig:tesslc} b), due to the Doppler boosting effect. These provide evidence that LC$_\BO$ arises from the ellipsoidal effect of a binary with an orbital period of 3.390497 days. In other words,  LC$_\BO$ and LC$_\BB$ are generated by two distinct binary systems.

\begin{figure*}[hptb!]
\centering
      \includegraphics[clip, width=0.9\textwidth]{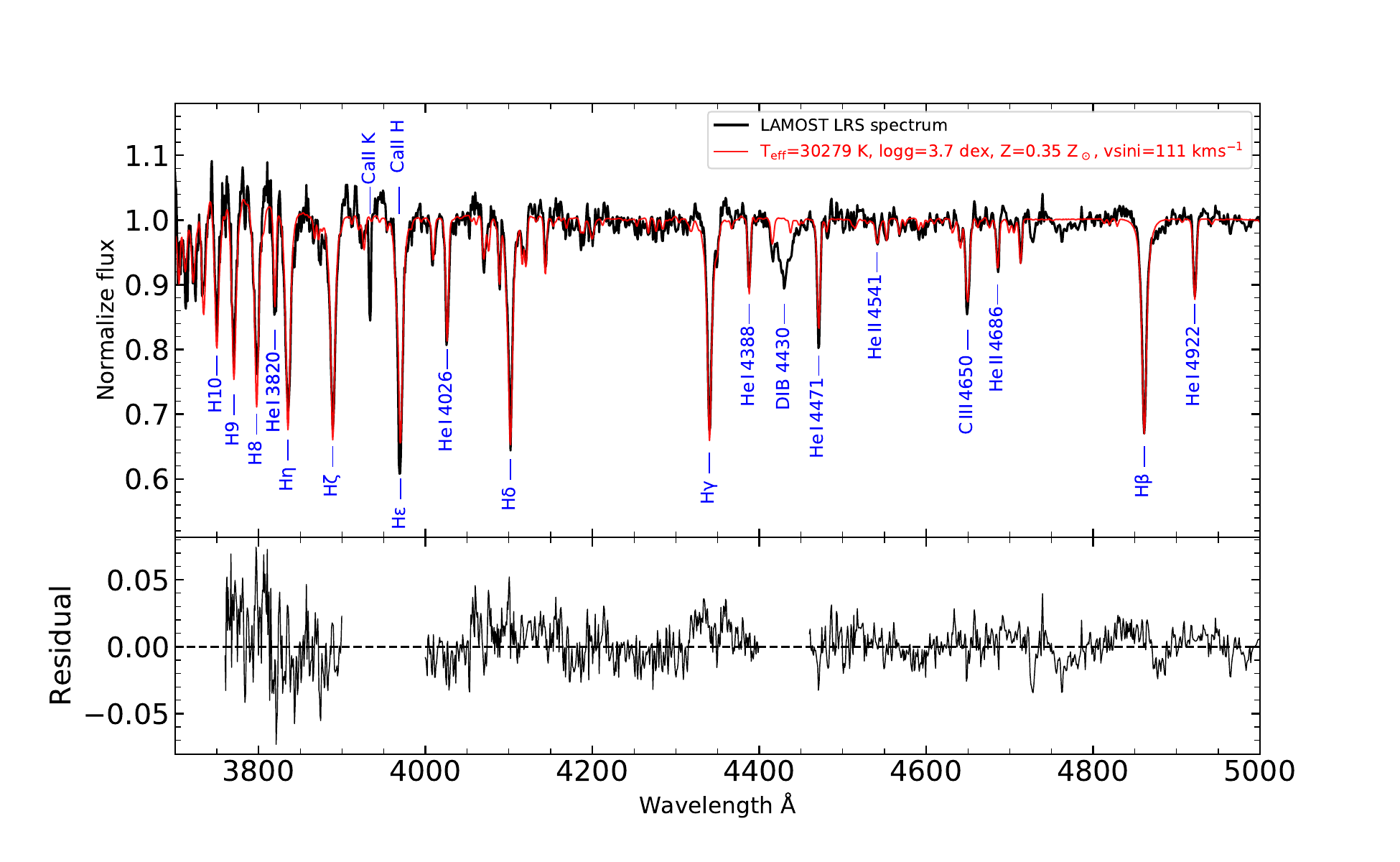}
\caption{The LAMOST-LRS spectrum. {\it Top}, the black line is observed spectra, the red line is the best-fit model of the MCMC approach; {\it Bottom}, the residual between the model and the observation spectrum, and the gaps in the spectrum are excluded when we determine the atmosphere parameters. The gaps are contaminated by the absorption lines of interstellar material (ISM): Ca II K, Ca II H, and diffuse interstellar bands (4430 $\Angstrom$).
}\label{fig:spec_lrs}
\end{figure*}

If the components of binary B$_\BO$ were bright enough and the spectrum resolution was greater than 2000, we would expect to observe double peaks in the CCF profiles when the spectra were observed at an orbital phase close to 0.25 or -0.25, given that the semi-amplitude of the radial velocity $K_\BOa$ is larger than 100 \kms. However, upon examining the CCF profiles of all spectra, we only found a single peak. Therefore, we use the LAMOST-LRS spectrum to measure the atmosphere parameters of star B$_\BOa$ and consider the contamination of the spectrum by the other three stars (details see Section~\ref{sec:BOa_atm}),
obtaining: $T_{\rm eff\BOa} = 30279\pm 1600$ K, $\log g_\BOa = 3.74 \pm 0.25$ dex, $Z_\BOa \sim 0.35 ~Z_\odot$, and $v\sin i_\BOa = 111 \pm 42$ ${\rm kms^{-1}}$. 
The exemplary spectroscopic fit is shown in Figure~\ref{fig:spec_lrs} (a). Moreover, the best-fitting model can also fit well with the spectrum of LAMOST-MRS (see Figure~\ref{fig:lamost-mrs}). 

\begin{figure*}[hptb!]
\centering
      \includegraphics[clip, width=0.9\textwidth]{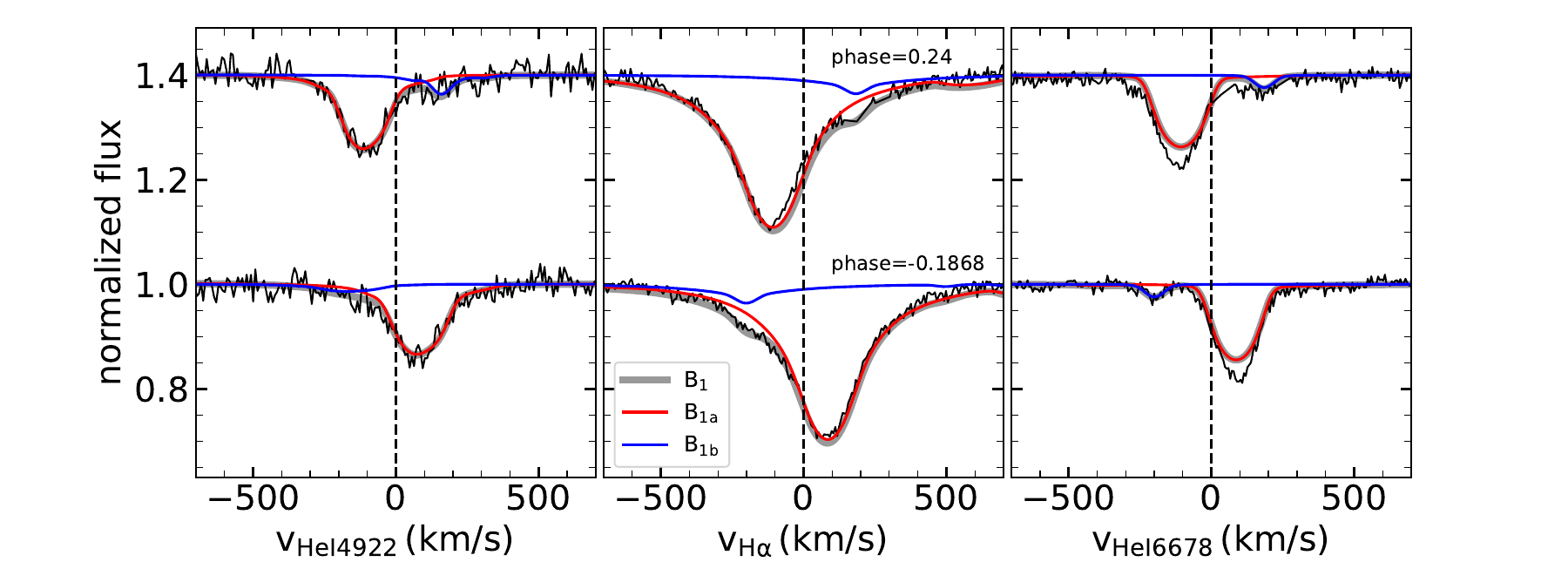} 
\caption{LAMOST-MRS spectra observed at the orbital phases of 0.24 and -0.1868 ($P=3.390497$ days) of binary B$_\BO$; the three panels are absorption lines of He I 4922, ${\rm H_\alpha}$ and He I 6678 from left to right; the black, red and blue thin lines are LAMOST-MRS spectra, synthetic spectra of B$_\BOa$, and B$_\BOb$; the gray thick lines are co-added synthetic lines of B$_\BOa$, and B$_\BOb$.
}\label{fig:spec_mrs_rv}
\end{figure*}

\subsection{Radial Velocity of Star B$_\BOb$ and Radius Ratio}\label{sec:fit_rv2}

We then carefully examined the LAMOST-MRS spectra taken at the orbital phases of -0.187 and 0.245 and discovered very faint lines that were moving in anti-phase with main absorption lines (such as He I 4922, ${\rm H\alpha}$, and He I 6678, as shown by the blue lines in Figure~\ref{fig:spec_mrs_rv}), which means the faint lines arise from the other component star B$_\BOb$. Furthermore, these lines displayed a higher velocity shift, indicating that star B$_\BOb$ has less mass than star B$_\BOa$. 

The star B$_\BOb{}$ should be a B-type star because the HeI and H${\rm \alpha}$ adsorption lines of B$_\BOb$ can be seen, as shown in Figure~\ref{fig:spec_mrs_rv}. The uncertainties of atmospheric parameters measured from LAMOST-MRS for B-type stars are very large. Additionally, the depths of these absorption lines in star B$_\BOb$ are comparable to the noise level, further complicating the determination of its atmospheric parameters.
Fortunately, its radial velocity can still be determined using the LAMOST-MRS spectra.


Its lower mass compared to star B$_\BOa$ implies that it is likely to be a main sequence star. Assuming specific atmosphere parameters, we could derive its radial velocity and radius ratio ($r_\BO = R_\BOb/R_\BOa$). For the sake of simplicity, we set the atmosphere parameters of B$_\BOa$ as fixed values: $T_{\rm eff\BOa} = 30279$ K, $\log g_\BOa = 3.74$ dex, $Z_\BOa=0.35~Z_\odot$. Since the radius ratio depends on atmosphere parameters of B$_\BOb$, we drive the radial velocity of B$_\BOb$ and radius ratio by exploring a grid of $T_{\rm eff\BOb}$ and $\log g_\BOb$, where $T_{\rm eff_\BOb}$ ranges from 19000 K to 29000 K with a step of 2000, and $\log g_\BOb$ ranges from 3.8 dex to 4.5 dex with a step of 0.1, $Z_{\BOb} = 0.35~Z_\odot$.

Therefore, the spectrum of binary B$_\BO$ depends on radial velocities ($v$), the projected rotation velocities ($v\sin i$) and radii ($R$) of its two components. It can be as a function of these parameters:   
\begin{equation}
\begin{split}
    &f(v_\BOa, v\sin i_\BOa, v_\BOb, v\sin i_\BOa, r) \\ 
    & = \frac{R_\BOa^2 f_\BOa(v_\BOa, v\sin i_\BOa)+R_\BOb^2 f_\BOb(v_{\BOb}, v\sin i_\BOb)}{d^2}\\
    &\propto f_\BOa(v_{\BOa}, v\sin i_{\BOa}) + r_\BO^2 f_\BOb(v_{\BOb}, v\sin i_{\BOb}),
\end{split}
\end{equation}
where $f_\BOa$ and $f_\BOb$ are synthetic spectra of stars B$_\BOa$ and B$_\BOb$ specific atomashpere paramters, respectively; $d$ is the distance of \thestar{}. We use an MCMC approach to fit the normalized spectra of LAMOST-MRS in wavelength ranges of [6520, 6600] and [6650, 6700] with the normalized broadened synthetic spectrum with a resolution of 7500 (the posterior probability distribution is similar to Eq. A1). This enables us to derive the radial velocities of star B$_\BOb$ and the radius ratio of star B$_\BOb$ to B$_\BOa$ across the grid of $T_{\rm eff\BOb}$ and $\log g_\BOb$. The radial velocities of B$_\BOb$ cannot be measured when the spectra were observed near phase 0, as the top panel of Figure~\ref{fig:tesslc} (c) shows that the number of solid triangles is five less than the number of open triangles. The radial velocities can be found in Table~\ref{tab:rv}. Moreover, the radial velocities were nearly independent of $T_{\rm eff \BOb}$ and $\log g_\BOb$.

\begin{figure}
    \centering
    \includegraphics[width=0.45\textwidth]{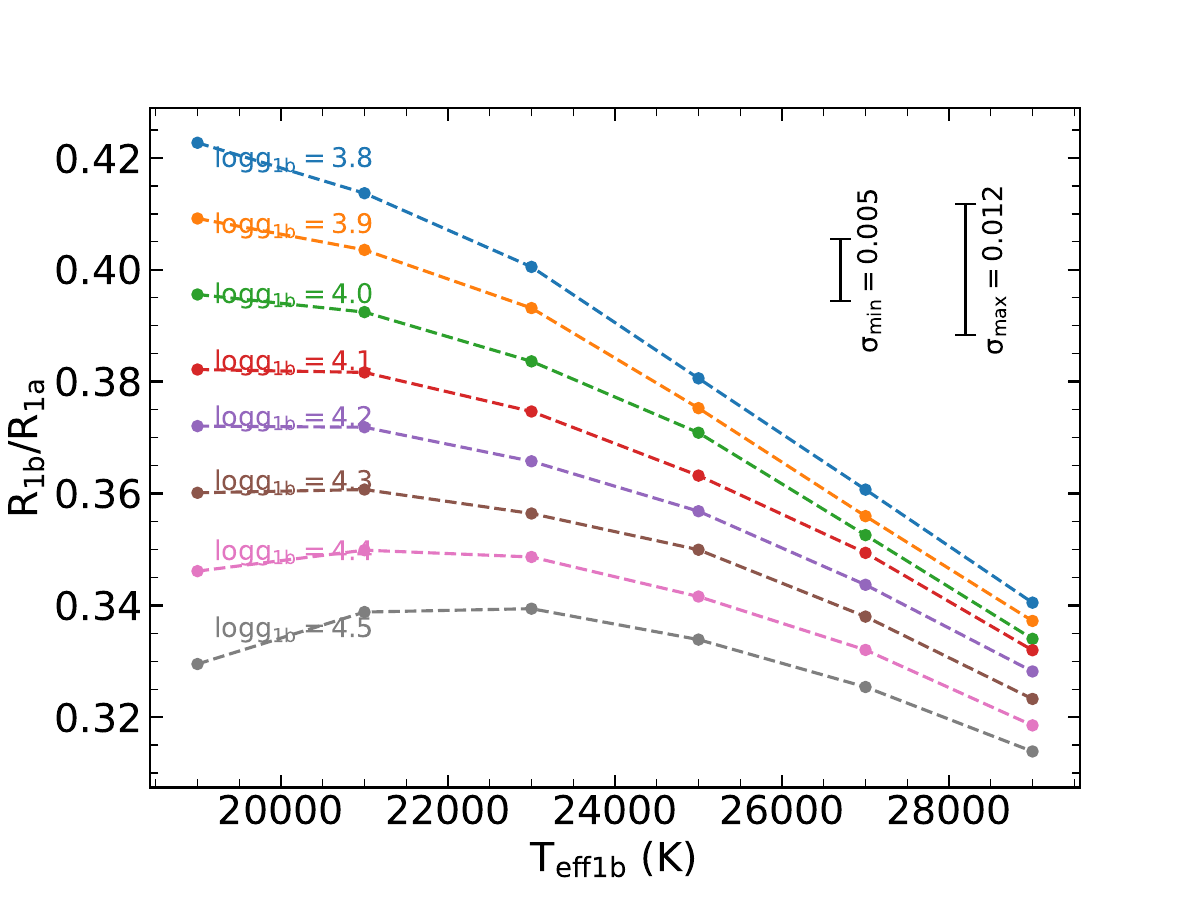}
    \caption{The radius ratio of star B$_\BOb$ to star B$_\BOa$ ($r_\BO = R_\BOb/R_\BOa$) varies with $T_{\rm eff\BOb}$ and $\log g_\BOb$. The vertical lines stand for the minimum and maximum errors.}
    \label{fig:R2tR1}
\end{figure}

In order to obtain an accurate radius ratio, we first selected 12 spectra that were observed near phases -0.25 and 0.25 (their locations in the folded radial velocity curve are indicated by the red triangles in the top panel of Figure~\ref{fig:tesslc} b). We then calculate the median values of the radius ratios of each point in $T_{\rm eff \BOb}$ and $\log g_\BOb$ grid, with their errors estimated by the bootstrap method using the package of \textit{Scipy}. The result is listed in Table~\ref{tab:R2tR1}. The radius ratio was found to decrease with an increase of $\log g_\BOb$. Furthermore, the radius ratio was observed to decrease with an increase of $T_{\rm eff\BOb}$, except when $\log g_\BOb$ was equal to 4.4 dex and 4.5 dex, as shown in Figure~\ref{fig:R2tR1}.

\subsection{Mass Ratio of Star B$_{1b}$ to B$_{1a}$}

Since we have measured the radial velocities of B$_\BOa$ and B$_\BOb$ at various time points, the mass ratio $q_\BO = m_\BOb m^{-1}_\BOa$ can be derived. We fit the radial velocities using the MCMC approach. The posterior distribution is
\begin{equation}\label{eq:likelihoodrv}
\begin{split}
     \ln[\mathcal{P}(\theta_{ v} \mid v_\BOa, v_\BOb)]&\propto \ln [\mathcal{P}(v_\BOa, v_\BOb \mid \theta_{\BO v})] \\
      &=\sum_{y=\BOa, \BOb}\ln[\mathcal{L}(v_y \mid \theta)]\\
      &=-\frac{1}{2}\sum_{y=\BOa, \BOb}\sum_{n=1}\frac{[v_{yn}-v_y(t_n; \theta_v)]^2}{\sigma^2_{v_{yn}}+s^2}\\
      &-\frac{1}{2}\sum_{y=\BOa, \BOb}\sum_{n=1}\ln [2\pi(\sigma_{v_{yn}}^2+s^2)],
\end{split}
\end{equation}
where $v_{yn}$ is the measured radial velocity of star B$_\BOa$ or B$_\BOb$, $\theta_v = (T_{\rm c\BO}, \sqrt{e_\BO}\cos\omega_\BOa, \sqrt{e_\BO}\sin\omega_\BOa,  K_\BOa, K_\BOb, v_{\gamma\BO}, s)$, $v_y(t_n; \theta_v)$ is the Kepler orbital velocity. (Note, $\omega_\BOb = \omega_\BOa + \pi$.)

We have determined $K_\BOa = 111.3\pm20.2$ \kms, $K_\BOb = 201.3\pm1.8$ \kms, and calculated $q =0.55\pm0.10$. The value of $K_\BOa$ is consistent with the result of $105.4\pm1.3$ \kms obtained in Section 3.1 within the error range, although it is slightly higher.

\begin{table*}
\centering
\begin{tabular}{c|ccc ccc cc}
\hline\hline
\diagbox[height=2\line, width=0.18\textwidth]{\parbox{1em}{$T_{\rm eff\BOa}$}}{\parbox{3em}{$\log g_\BOa$}} & 3.8 & 3.9 & 4.0 & 4.1 & 4.2 & 4.3&4.4& 4.5 \\ 
\hline
19000&0.423(9)&0.409(9)&0.396(9)&0.382(11)&0.372(12)&0.360(11)&0.346(11)&0.329(12)\\
21000&0.414(8)&0.404(8)&0.392(8)&0.382(8)&0.372(8)&0.361(9)&0.350(9)&0.339(9)\\
23000&0.401(7)&0.393(7)&0.384(8)&0.375(7)&0.366(7)&0.356(7)&0.349(7)&0.339(8)\\
25000&0.381(6)&0.375(6)&0.371(7)&0.363(7)&0.357(7)&0.350(7)&0.342(7)&0.334(6)\\
27000&0.361(6)&0.356(6)&0.353(6)&0.349(7)&0.344(6)&0.338(7)&0.332(7)&0.325(7)\\
29000&0.340(6)&0.337(6)&0.334(5)&0.332(6)&0.328(6)&0.323(6)&0.319(6)&0.314(6)\\
\hline
\end{tabular}
\caption{The grid of radius ratio of star B$_\BOb$ to star B$_\BOa$ ($r_\BO = R_\BOb/R_\BOa$) varied with $T_{\rm eff\BOb}$ and $\log g_\BOb$.}
\label{tab:R2tR1}
\end{table*}

\section{evidence of quadruplicity}\label{sec:quadruple}
Since the pixel size of TESS is 21$''$, the TESS light curve might be contaminated by its vicinity stars. So we extract a light curve from a single pixel from the full-frame image (FFI) centered at \TYC\ by \texttt{Tesscut} \citep{Brasseure2019ascl}, the ellipsoidal and eclipse features still exist. 

It's worth noting that we initially presumed \TYC\ to be a binary system (B$_\BO$), and the eclipsing light curve originated from a nearby eclipsing binary B$_\BB$. We inferred the minimum luminosity of binary B$_\BB$ by comparing the eclipse depth of LC$_\BB$ with the $G\RP = 10.12$ mag magnitude of \thestar{} (for more details, refer to Section~\ref{sec:vicinity_eclipse}).

To validate this assumption, we searched for light curves of stars within a 60$''$ angular radius in the Zwicky Transient Facility (ZTF: \citealt{Masci2019PASP, Bellm2019PASP}) Data Release (DR9). Specifically, we searched for eclipsing binaries with a period of $P_\BB = 2.43781$ days within the magnitude range $G\RP \in [10.12, 15.12]$ mag. However, none was found.
TYC 3340-2437-1, therefore, might be a quadruple system composed of binary B$_\BO$ and binary B$_\BB$.


Moreover, binary B$_\BB$ should be fainter than star B$_\BOb$, since no signal of binary B$_\BB$ can be detected from the LAMOST-MRS spectra (see Figure~\ref{fig:spec_mrs_rv}). We assume that the distribution of nearby stars around \TYC{} is uniform in terms of solid angle. Counting from {\it Gaia} DR3, there are 1695 stars located in the radius of $0.5^\circ$ centered at \TYC{} with $G_{\rm RP}$ from 10.12 to 15.12 mag, so the steradian density is $\rho = 7.08 \pm 0.17\times 10^{6}$ ${\rm sr^{-1}}$. The probability of optical double $\mathcal{P} \sim \rho \pi R_{\rm ang}^2 = 8.4\pm0.2\times 10^{-5}$ adopted effective angular resolution of {\it Gaia} $R_{\rm ang} \sim 0.4" = 1.94\times 10^{-6} $ rad\footnote{\url{https://www.cosmos.esa.int/web/gaia/dr2}}. In other words, \thestar{} has more than 99.99\% probability to be a 2+2 quadruple system. 

This is reinforced by the fact that the parameters of astrometric excess noise $\epsilon = 0.1$ mas and significance of the noise $D = 19.68$ in {\it Gaia} DR3, because $\epsilon > 0$ mas and $D > 2$ represent an additional intrinsic scatter term in the astrometric solution \citep{Lindegren2012A&A}. In this case, it might be caused by orbital motion between binary B$_\BO$ and binary B$_\BB$.


\subsection{Characterizing the System}

Combining the binary mass function of star B$_{\BO}$
\begin{equation}
    f_{\BOb} = \frac{m^{3}_{\BOb}\sin^{3}i_{\BO}}{(m_{\BOa}+m_{\BOb})^{2}} = \frac{P_{\BO}K^{3}_{\BOa}}{2\pi G}(1-e_\BO^{2})^{3/2}
    \label{eq:massfunc}
\end{equation}
and $g_{\BOa} = Gm_\BOa R_\BOa^{-2}$, we can obtain
\begin{equation}
    R_\BOa^2\sin^3i_\BOa = \frac{P_{\BO}K^{3}_{\BOa}}{2\pi q_\BO^3 g_\BOa}(1-e_\BO^{2})^{3/2}(1+q_\BO)^2.
    \label{eq:Ri_rv}
\end{equation}
Since the variables on the right-hand side of Eq. (\ref{eq:Ri_rv}) were obtained from LAMOST-LRS spectra and the radial velocity curve of binary B$_\BO$ (for details, see Section~\ref{sec:rv}), and $R_\BOa^2\sin^3i_\BOa$ remains constant.

The characteristics of an ellipsoidal light curve are predominantly influenced by the mass ratio, orbit inclination, and the ratios of star radii to the semi-major axis (for details, see Section 3.2.1 of \citealt{PHOEBE2018maeb.book.....P}). For \thestar, LC$_\BO$  is primarily a result of the ellipsoidal variations of star B$_\BOa$, given that $R_\BOb < 0.5 R_\BOa$ (see Figure~\ref{fig:R2tR1}) and $q_\BO\sim 0.5$. It depends mainly on $q_\BO$, $i_\BO$, $R_\BOa$, and the smi-major axis $A_\BO$ of binary $B_\BO$. Notably, combining $A_\BO^3 = Gm_\BOa(1+ q_\BO)P_\BO^2(2\pi)^{-2}$ and $g_{\BOa} = Gm_\BOa R_\BOa^{-2}$ indicates that $A_\BO$ is solely determined by $R_\BOa$ in this particular case. Therefore, LC$_\BO$ is expressed as a function of $i_\BOa$ and $R_\BOa$.

Considering that the luminosity ratio of binary B$_\BB$ to B$_\BO$ is less than 0.02 (see Section~\ref{sec:vicinity_eclipse}), it can be ignored that LC$_\BO$ is influenced by the light coming from binary B$_\BB$. With the two variables and the two functions, we can constrain both $i_\BO$ and $R_\BOa$ by a simultaneous fitting of the ellipsoidal curve and the radial velocity curve. This allows us to calculate the masses of stars B$_\BOa$ and B$_\BOb$.

Based on the atmosphere parameters ($T_{\rm eff\BOa} = 30249$ K, $\log g_\BOa = 3.74$ dex, and $Z_\BOa = 0.35$ $Z_{\odot}$) of star B$_\BOa$, it appears to be a main sequence star. Additionally, we do not detect any signal of star B$_{\BBa}$ and star B$_{\BBb}$ from our current spectra, indicating that their luminosities are less than that of star B$_{\BOb}$. So, we assume that star B$_\BOb$, B$_\BBa$, and B$_\BBb$ are also main sequence stars. The luminosity of star B$_{\BOb}$ can be obtained according to the relation between the luminosity and mass of the main sequence star, given a specific mass. We use isochrones of PAdova and Trieste Stellar Evolution Code (PARSEC) stellar model \citep{Bressan2012MNRAS, Chenyang2014MNRAS} to interpolate the luminosity from the mass of star B$_\BOb$. The masses of star B$_\BBa$ and star B$_\BBb$ can then be interpolated based on their luminosities, which can be estimated by the relative eclipse depths of binary B$_\BB$ (see Section~\ref{sec:vicinity_eclipse} and  Figure~\ref{fig:tesslc} (c) for clarity).

Consequently, the semi-major axis of binary B$_\BB$ can be calculated with its masses and period. Additionally, the eclipse widths of LC$_\BB$ are influenced by the radii of star B$_\BBa$ and star B$_\BBb$ (for details, refer to Section 3 of \citealt{PHOEBE2018maeb.book.....P}). Furthermore, their luminosities would increase as the inclination of binary B$_\BB$ ($i_\BB$) decreases. For example, given a luminosity ratio of star B$_\BBb$ to star B$_\BBa$, if $i_\BB=90\degree$, the primary eclipse of the binary would be the deepest, and the least luminosity of the star is needed to reach the relative depth. By simultaneously fitting the eclipse widths of LC$_\BB$ with their masses and orbital period, we can determine their radii and inclination of binary B$_\BB$.

We assumed that the other three stars have the same metallicity as star B$_\BOa$ and fixed orbital periods of binary B$_\BO$ (3.390497 days) and binary B$_\BB$ (2.43781 days), the transit time of binary B$_\BB$ (2458817.49056 BJD), and the effective temperature (30279 K) of star B$_\BOa$ to simplify the analysis. Then an MCMC approach is performed to fit the TESS LC and radial velocity curves of binary B$_\BO$ with these assumptions. The posterior distribution can be obtained using these priors; for details, see Section~\ref{sec:mcmclcrv}. The best-fit model is shown in Figure~\ref{fig:tesslc} (a). We obtain that the masses of \thestar{} are ($11.47_{-0.26}^{+0.28}$ + $5.79_{-0.12}^{+0.13}$) + ($5.20_{-0.69}^{+0.38}$ + $2.02_{-0.08}^{+0.16}$) $M_{\odot}$. The inclinations of binary B$_\BO$ and B$_\BB$ are $55.94^{+0.55}_{-0.49}$ and $78.2^{+3.16}_{-0.96}$ degrees. The other parameters can be found in Table~\ref{tab:para} and Figure~\ref{fig:corner_lcrv}. 

We noted that $T_{\rm eff2a}$, $\log g_{\rm 2a}$ and $R_{\rm 2a}$ are similar to the corresponding parameters of star B$_\BOb$. However, despite these similarities, we did not detect any discernible signal in the LAMOST-MRS spectrum corresponding to B$_\BBa$. This lack of detection could be attributed to two reasons. Firstly, B$_\BBa$ doesn't exhibit a noticeable anti-phase movement with respect to star B$_\BOa$, making it challenging to distinguish its spectral lines since the depths of the absorption lines of B$_\BBa$ are comparable to (or lower than) the noise level in our observations. It is also plausible that the luminosity of B$_\BBa$ is overestimated, although this cannot be conclusively determined with our spectra.

From the bottom panel of Figure~\ref{fig:tesslc} (a), we can see that there are some residuals with absolute values exceeding three times their errors, see bottom panel of Figure~\ref{fig:tesslc} (a). When we fold the residuals into orbital periods of binary B$_\BO$ and binary B$_\BB$, we do not observe any period signals, as shown in the bottom panels of Figure~\ref{fig:tesslc} (b) and (c). This indicates that these larger residuals are not generated by the orbital motions of binary B$_\BO$ or binary B$_\BB$. There are several possible reasons for this, such as 1) the de-trending process of the light curve; 2) pulsations of the four components; 3) the orbital motion between binary B$_\BO$ and B$_\BB$; 4) underestimated errors of the light curve. The small errors in the derived parameters (see Figure~\ref{fig:corner_lcrv}) may suggest the light curve errors were underestimated. However, we cannot figure out the exact cause of the large residual with the current observation data.

\begin{table*}[htpb!]
    \centering
    \begin{tabular}{c|cc|cc}
    \hline\hline
    Parameter & B$_\BOa$ & B$_\BOb$ & B$_\BBa$ & B$_\BBb$\\
    \hline
    $T_{\rm eff}$ (K) & $30279\pm 1600$ [fixed] & $19299_{-215}^{+374}$ & $18023_{-2516}^{+997}$ & $8140_{-497}^{+1492}$ \\
	$\log g$ (${\rm cm\, s^{-2}}$)  & $3.717_{-0.003}^{+0.003}$ & $4.271_{-0.029}^{+0.029}$ & $4.256_{-0.099}^{+0.039}$ & $3.955_{-0.068}^{+0.244}$ \\
	$m$ ($M_{\odot}$)  & $11.47_{-0.26}^{+0.28}$ & $5.79_{-0.12}^{+0.13}$ & $5.20_{-0.69}^{+0.38}$ & $2.02_{-0.08}^{+0.16}$ \\
	$R$ ($R_\odot$)  & $7.77_{-0.09}^{+0.09}$ & $2.92_{-0.11}^{+0.12}$ & $2.83_{-0.11}^{+0.14}$ & $2.49_{-0.57}^{+0.17}$ \\
	$R_L$ ($R_{\odot}$)  & $10.78_{-0.09}^{+0.09}$ & $7.89_{-0.06}^{+0.06}$ & $6.84_{-0.47}^{+0.20}$ & $4.43_{-0.06}^{+0.12}$ \\
	$\log L$ ($L_{\odot}$)  & $4.66_{-0.01}^{+0.01}$ & $3.04_{-0.04}^{+0.03}$ & $2.87_{-0.22}^{+0.11}$ & $1.39_{-0.07}^{+0.12}$ \\
$Z$ ($Z_\odot$)&$\sim0.35$&--&--&--\\
	$K$ (\kms)  &  $101.9_{-1.0}^{+1.1}$ & $201.7_{-1.6}^{+1.7}$ & -- & -- \\
    \hline
    $P$ (day)&\multicolumn{2}{c|}{3.390497(30) [fixed]} & \multicolumn{2}{c}{2.43781(16) [fixed]}\\
    $T_{\rm c}$ (BJD)&\multicolumn{2}{c|}{ $2458821.51571(86)$} & \multicolumn{2}{c}{2458817.49056(75) [fixed]}\\
    $v_\gamma$ (\kms)&\multicolumn{2}{c|}{$-14.42_{-0.77}^{+0.76}$} & \multicolumn{2}{c}{--}\\
    $i$ (\degree)&\multicolumn{2}{c|}{$55.94_{-0.49}^{+0.55}$} & \multicolumn{2}{c}{$78.20_{-0.96}^{+3.16}$}\\
    $A$ ($R_{\odot}$)&\multicolumn{2}{c|}{$24.55_{-0.17}^{+0.19}$} & \multicolumn{2}{c}{$14.76_{-0.41}^{+0.27}$}\\
    $e$ &\multicolumn{2}{c|}{0.01101(59)} & \multicolumn{2}{c}{0 (fixed)}\\
    $q$ &\multicolumn{2}{c|}{0.5052(72)} & \multicolumn{2}{c}{$0.385_{-0.025}^{+0.093}$}\\
    $\omega$ ($\degree$) &\multicolumn{2}{c|}{$142.2\pm3.6$} & \multicolumn{2}{c}{--}\\
    \hline
    $m_{\rm total}$ ($M_{\odot}$)&\multicolumn{4}{c}{$26.85_{-0.57}^{+0.62}$}\\
    $\varpi$ (mas)&\multicolumn{4}{c}{$0.332- (-0.033)\pm0.018$}\\
    RUWE &\multicolumn{4}{c}{1.1073}\\
    $\epsilon$ (mas) &\multicolumn{4}{c}{0.10}\\
    $D$ &\multicolumn{4}{c}{19.68}\\
    ($\alpha$, $\delta$) ($\degree$)&\multicolumn{4}{c}{( 62.79471725, +50.70821829)}\\
    ($l$, $b$) ($\degree$)&\multicolumn{4}{c}{(151.92316895, -0.56520738)}\\
    $(\mu_\alpha,\, \mu_\delta)$ (${\rm mas\, yr^{-1}}$) & \multicolumn{4}{c}{($-0.954\pm0.019$, $-1.176\pm0.014$)}\\
    ($G$, $G_{\rm BP}$, $G_{\rm RP}$) (mag) &\multicolumn{4}{c}{(10.91, 11.55, 10.12)}\\
    \hline
    \end{tabular}
    \caption{Stellar, orbital, and astrometric parameters of \TYC. The stellar and orbital parameters are derived using an MCMC approach to fit the radial velocity curve of binary B$_\BO$ and the light curve of the quadruple system. Astrometric parameters are from {\it Gaia} EDR3 \citep{Gaiadr32022A&A}, with a zero-point correction applied to the parallax \citep{Lindegren2021A&A}. Quantities shown are effective temperature $T_{\rm eff}$, surface gravity $\log g$, stellar mass $m$, radius $R$, Roche lobe radius $R_L$, luminosity $\log L$, metallicity $Z$, the semi-amplitude of radial velocity $K$, the orbital period $P$, the zero-point of the ephemeris $T_c$, the systemic velocity of the binary $v_{\gamma}$, the orbital inclination $i$, the smi-major axis $A$, the orbital eccentricity $e$, the mass ratio $q$, the longitude of periastron $\omega$, the total mass of the quadruple system $m_{\rm total}$, parallax $\varpi$, the Renormalised Unit Weight Error RUWE, the astrometric excess noise $\epsilon$ (astrometric\_excess\_noise), the significance of astrometric excess noise $D$ (astrometric\_excess\_noise\_sig), the ICRS coordinates ($\alpha, \delta)$,  the Galactic coordinates ($l, b$), proper motions in the right ascension $\mu_\alpha$, and declination $\mu_\delta$, the magnitude of {\it Gaia} ($G$, $G_{\rm BP}$, $G_{\rm RP}$). }
    \label{tab:para}
\end{table*}


\section{Distance from SED fitting}
\label{sec:sed_distance}
The SED fitting was performed following a strategy similar to the one used in the Python package \href{https://speedyfit.readthedocs.io/en/stable/index.html}{Speedyfit}. We prepared grids of fluxes passing through filters of {\it Gaia}, APASS, 2MASS, and allWISE using SEDs of TLUSTY OSTAR2002~\citep{Lanz2003ApJS}, TLUSTY BSTAR2006~\citep{Lanz2007ApJS..169...83L}, and Kurucz~\citep{Castelli2003IAUS..210P.A20C}.

\begin{figure}[htpb!]
    \centering
    \includegraphics[width=0.47\textwidth]{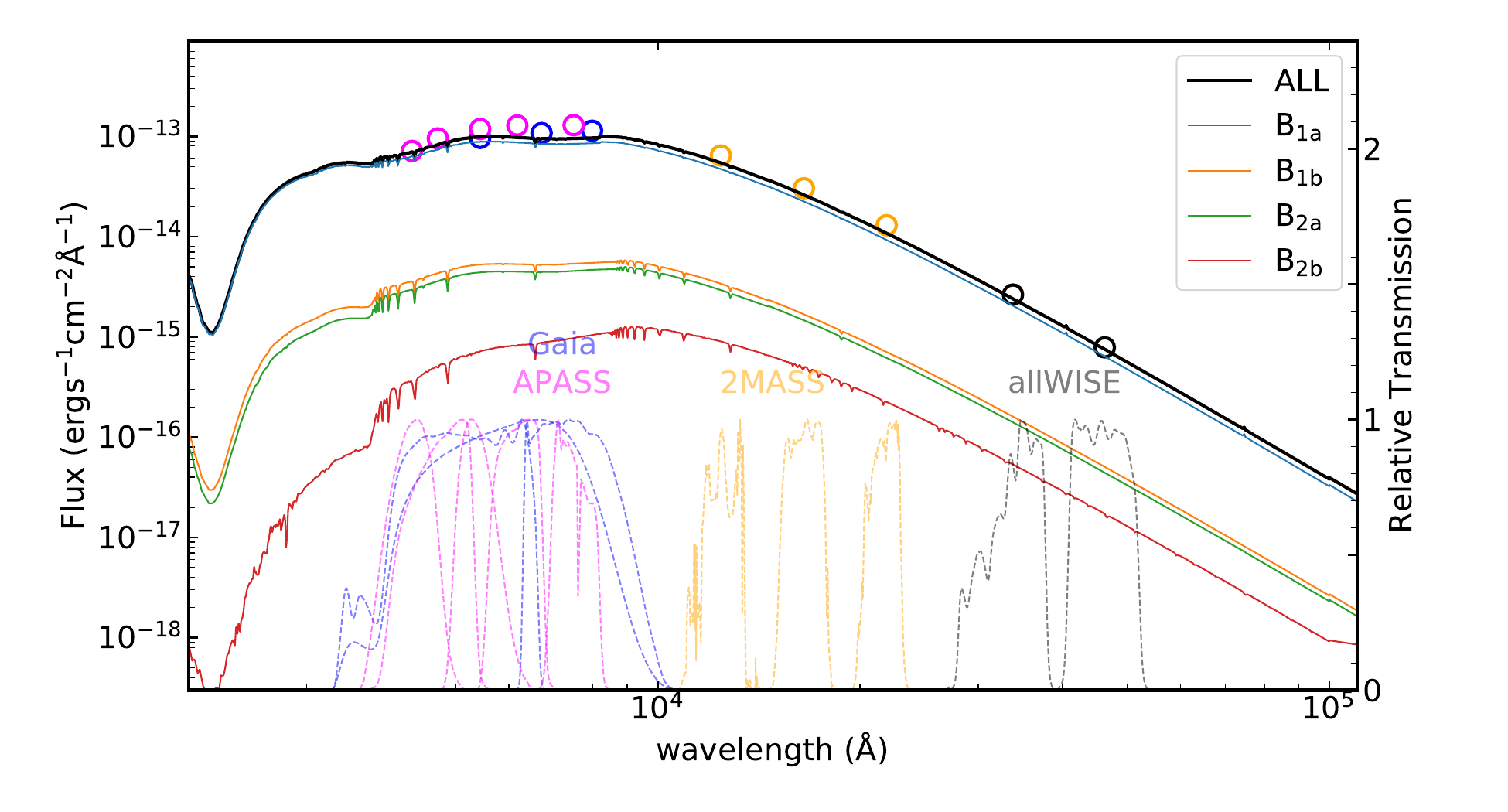}
    \caption{Comparson of synthetic and observed photometry.
    {\it Left axis} is the spectral energy distribution; the open dots represent the fluxes from {\it Gaia} EDR3 $G_{\rm BP}$-, $G$- and $G_{\rm RP}$-bands, APASS $B$-, $V$-, $G$-, $R$-, $I$-bands, 2MASS $J$-, $H$- and $K_{\rm s}$-bands, and ALLWISE $W_1$-, $W_2$-bands, which are downloaded from \href{http://cdsportal.u-strasbg.fr/}{CDS portal}; the black curve stands the best-fitting SED of \thestar{} summed by SEDs of stars B$_\BOa$, B$_\BOb$, B$_\BBa$, and B$_\BBb$.
    {\it Right axis} shows the transmission curve normalized at the maximum. The transmission curves are labeled as the same colors of open dots.}
    \label{fig:sedfit}
\end{figure}

Fixing the effective temperatures (30279, 19299, 18023 and 8140 K) and surface gravities (3.717, 4.271, 4.256 and 3.955 dex) of the four components of \thestar{} derived from the light curve and radial velocity curve fitting, we interpolated fluxes for stars B$_{\BOa}$, B$_{\BOb}$, B$_{\BBa}$, and B$_{\BBb}$ from the flux grids of TLUSTY OSTAR2002, TLUSTY BSTAR2006, TLUSTY BSTAR2006, and  Kurucz, respectively. We assumed that their radii followed Gaussian priors with means of 7.77, 2.92, 2.83, and 2.02, and standard deviations of 0.09, 0.12, 0.14, and 0.57 $R_\odot$. We used an MCMC approach to fit the observed fluxes of {\it Gaia} EDR3 $G_{\rm BP}$-, $G$-, and $G_{\rm RP}$-bands, APASS $B$-, $V$-, $G$-, $R$-, and $I$-bands, 2MASS $J$-, $H$- and $K_{\rm s}$-bands, and ALLWISE $W_1$- and $W_2$-bands. The best-fit model is shown in Figure~\ref{fig:sedfit}.

Our analysis yielded a distance of $2.31\pm0.03$ kpc and a color excess of $E(B-V) =1.198\pm0.002$ mag for \thestar{}. This distance is less than the one ($d_{\it Gaia}=2.75\pm0.14$ kpc) derived from {\it Gaia} DR3 parallax. This discrepancy indicates that the astrometric solution of \thestar{} cannot be adequately explained by the single-star model again.

The angle distance between star B$_\BOa$ and star B$_\BOb$ (star B$_\BBa$ and star B$_\BBb$) is about 49.4 $\mu{\rm as}$ (29.7 ${\rm\mu as}$). This angle distance is too small to have a significant impact on the astrometric solution for {\it Gaia}. Therefore, the solution of {\it Gaia} parallax might be influenced by the orbital motion between binary B$_{\BO}$ and binary B$_{\BB}$.

\section{ENVIRONMENT}\label{sec:cloud}

\thestar{} is a star in the SBN and identified by using the Wide Field Infrared Explorer (WISE: \citealt{Wright2010AJ}) image (see the inset panel of Figure~\ref{fig:comap}). It indicates that \thestar{} is surrounded by ISM. We calculate the radial velocities of Na D1 and Na D2 absorption lines observed by LAMOST-LRS and (B)YFOSC (see Table~\ref{tab:rv}) and obtain $v_{\rm Na} = -18.5_{-11.8}^{+6.9}$ \kms. Since we know the Galactic coordinates and the distance ($d = 2.31\pm0.03$ kpc derived from SED fit) of the star, we calculate the radial velocity of the clouds where the star was born $v_{\rm ev}\sim -18.6$ \kms{} with respect to the local standard of rest (LSR) by assuming that the Sun’s motion with respect to the LSR is (11.1, 12.24, 7.25) \kms and the circular speed is 238 \kms~\citep{Schonrich2010MNRAS, Schonrich2012MNRAS}. It suggests that the comoving CO cloud of the star has a radial velocity similar to the ISM.

We download the $\rm ^ {12}CO (J=1-0)$ data around \TYC{} from the Milky Way Imaging Scroll Painting (MWISP) project \citep{Suyang2019ApJS} and integrate emission of $\rm ^ {12}CO (J=1-0)$ from -30.3 to -11.7 \kms, which are the 16th and 84th percentile of the radial velocities of the Na D1 and Na D2 absorption lines. The radial velocity is also consistent with the systematic radial velocity of binary B$_\BO$ of -14.4 \kms. As shown in Figure~\ref{fig:comap}, \thestar{} is located in a region with $^{12}$CO integrated intensity of $\sim 20$ K \kms. The cloud might be the reservoir of star formation to generate the quadruple system.

\section{Discussion}

If it were true that the {\it Gaia} DR3 astrometric solution of \thestar{} was affected by the orbital motion between binaries B$_{\BO}$ and B$_{\BB}$, then the outer orbital period $P_{\rm out}$ would be comparable to the coverage time of the astrometric time series. Considering that the {\it Gaia} DR3 catalog was created from raw data collected by the {\it Gaia} instruments during the first 34 months \citep{Gaiadr32022A&A}, the outer period could be a few years or a decade.

Assuming $P_{\rm out}$ is 5 (or 10) years and it is a circular orbital, the outer semi-major axis $a_{\rm out} \sim 1826 (2899)$ $R_\odot$ between binary B$_\BO$ and B$_\BB$. Following the Equation (12) of \cite{Georgakarakos2008CeMDA}, we obtain the largest stable criteria $(a_{\rm out}/a_{\rm in})_{\rm c} \sim 7$, assuming binary B$_\BO$ as the inner binary, B$_\BB$ as the third star with a mass of ($m_\BBa + m_\BBb$), and the third star exhibiting prograde motion. Therefore, the quadruple system would be stable due to $a_{\rm out}/a_{\rm in} \sim 74 > 7$.

The orbital velocities of binary B$_\BO$ and B$_\BB$ would be 14.9 (or 11.8) and 35.7 (or 28.3) \kms. If the outer orbit had an eccentricity $e_{\rm out} > 0$, the orbital velocity of binary B$_\BO$ at the periapsis would be greater than 14.9 \kms{} and even greater than 20 \kms{}. This raises the probability that the arc-like feature of the bow-shock nebula may result from the interaction between the surrounding ISM and stellar wind expelled when binary B$_\BO$ near periapsis.

\begin{figure}
    \centering
    \plotone{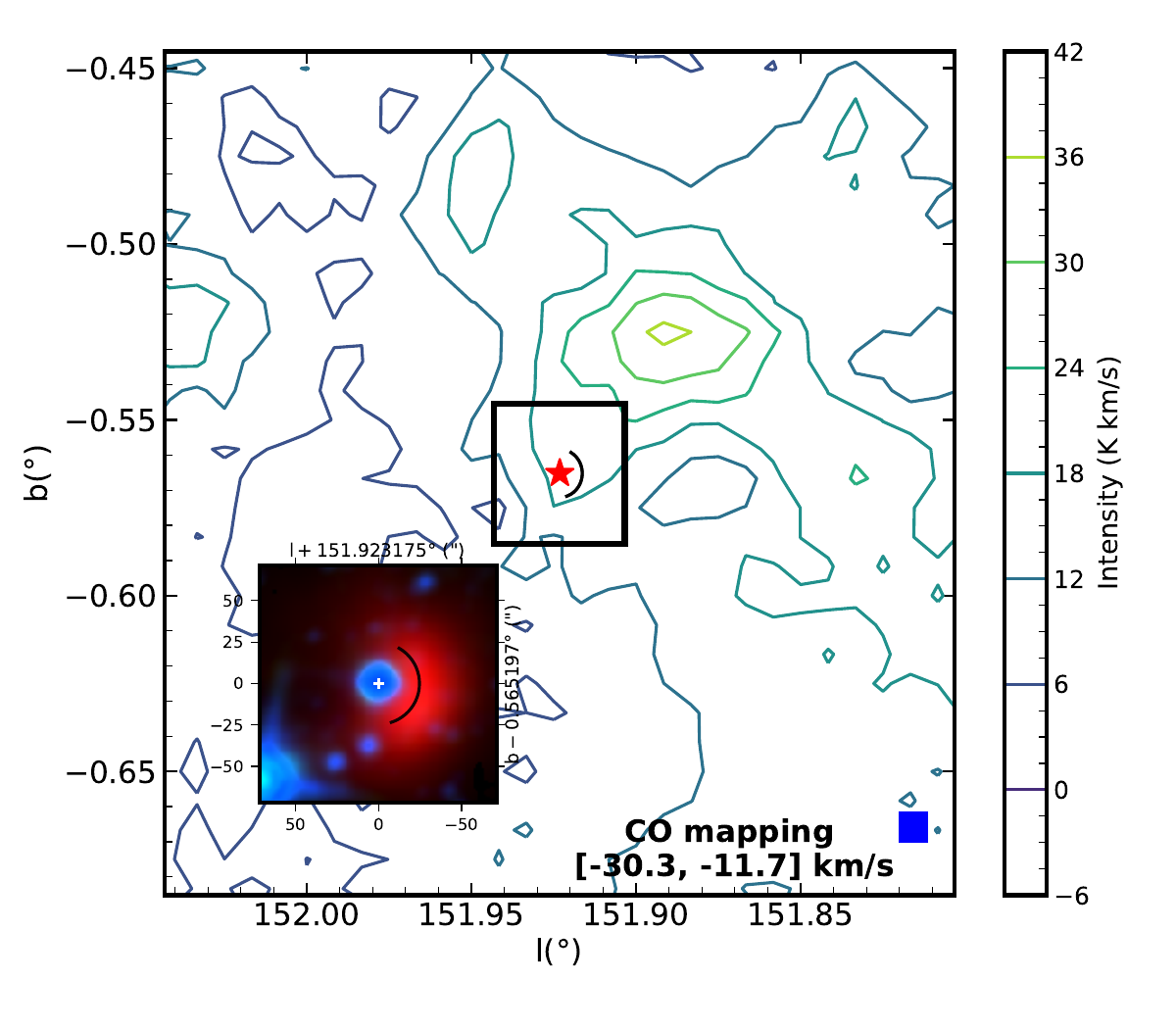}
    \caption{Integrated emission of $\rm ^ {12}CO (J=1-0)$ of MWISP in the interval of [-30.3, -11.7] ${\rm kms^{-1}}$, which is the 68\% confidence interval of the radial velocity measured by Na D1 and Na D2 absorption lines. The blue square is the pixel size of MWISP image.
    {\it Insert panel} is WISE image colored by bands of W1 (blue), W2 (green), and W4 (red). It is the same size as the black quadrangle of the main panel. The black curves (arc) of the main and insert panels have the same radius of $\sim 24.6"$ (0.33 pc), which go through the intensity peak of WISE W4 and indicate the position of the SBN.}
    \label{fig:comap}
\end{figure}

In such a scenario, one would expect the bow-shock nebula to exhibit discrete features in the direction of the arc-like structure, as these shocks could form at different times when binary B$_\BO$ approaches periapsis. However, no such discrete features are evident in the WISE image (see insert panel of Figure~\ref{fig:comap}). This absence could be attributed to limitations in the spatial resolution of the WISE telescope. Alternatively, the bow-shock nebula might have formed from the unitary motion of the quadruple system, similar to the bow-shock nebula of a runaway star. It is also plausible that the bow-shock nebula is an ``in situ'' feature, supported by interactions between \thestar{} and an outflow of hot gas from a nearby star-forming region or the H II region (the bow-shock theory can be found from \cite{Kobulnicky2016} and the references therein).

If $P_{\rm out}$ were 5 (or 10) years, the semi-major axis would be about 8.5 (or 13) au. This is significantly less than 500 au, suggesting that \thestar{} might have formed through disk fragmentation \citep{Offner2022arXiv}. However, it's noteworthy that 
the orbital inclinations of binaries B$_\BO$ and B$_\BB$ are $55.9^{+0.55}_{-0.49}$ and $78.20^{+3.16}_{-0.96}$ degrees, respectively. These inclinations indicate that the orbital orientations of the inner binaries are not coplanar, which contrasts with the idea that disk fragmentation mechanisms typically result in multiple systems with initially similar orientations.

However, if the eccentricity of the outer orbit were greater than zero, an interesting possibility arises. The Kozai–Lidov mechanism can induce large-amplitude oscillations of eccentricities and inclinations for highly inclined triple systems (\citealt{Naoz2013MNRAS.431.2155N} and references therein). In a similar context, it's conceivable that binaries B$_\BO$ and B$_\BB$ were initially formed with similar orientations, but with non-zero eccentric of outer obit, which would lead to changes in their orientations over time through secular evolution of the system. This issue might be figured out when the epoch positions of {\it Gaia} are released in the future.

\section{conclusion}

Cross-matching TESS LC with the O-type stars found by LAMOST, we find a hierarchical quadruple system \thestar. It is excluded to be an optical double star and has a probability of more than 99.99\% to be a (2+2) quadruple system by analyzing the data of {\it Gaia} DR3 and ZTF. Its inner orbital periods are 3.390497(30) days and 2.43781(16) days. Assuming the four stars are main sequence stars, the total mass of the quadruple system is estimated to be about $24.48= (11.47 + 5.79) + (5.2 + 2.02)$ $M_\odot$ by a simultaneous fitting of both light curve and radial velocity curve.

The line-of-sight inclinations of the inner binaries are estimated $i_\BO = 55.94_{-0.49}^{+0.55}$ and $i_\BB=78.20_{-0.96}^{+3.16}$. \thestar{} is a quadruple system in SBN and surrounded by cloud with $^{12}$CO integrated intensity of $\sim 20$ K ${\rm km\,s^{-1}}$ which could be the reservoir for forming \thestar. The outer orbital period of \thestar{} might be a few years or a decade. The quadruple system might formed through the disk fragmentation mechanism. The arc-like feature of SBN and non-co-planar of the inner binary might be due to an eccentric outer orbit. The outer orbit period and outer eccentric could potentially be determined when the epoch positions of {\it Gaia} are released in the future, which can also be used to constrain the mechanism of star formation.

\section{acknowledgments}


The authors thank Hao Tian, Man I Lam, Jie Lin, Sheng-Hong Gu, Fang Yang, Er-Lin Qiao, Jian-Rong Shi, Hai-Liang Chen, Zheng-Wei Liu, Xiang-Cun Meng, Yan Gao, Bo Wang, Shu Wang, Xiao Zhou, Zhi-Cun Liu, Yan-Jun Guo and Jian-Ping Xiong for valuable discussion; and thank Jin-Ming Bai, Yu-Feng Fan, Xiao-Guang Yu, Jian-Guo Wang, Kai-Xing Lu, Ju-Jia Zhang, Jie Zheng, Hong Wu for the suggestion and assistance of observation. We acknowledge the support of the staff of the Lijiang 2.4m telescope and the Xinglong 2.16m telescope. We thank the anonymous reviewer for the valuable suggestions and comments.

This work is supported by the National Key R\&D Program of China grant (Nos. 2021YFA1600401, 2021YFA1600400, 
and 2019YFA0405501.L.C.), 
the Natural Science Foundation of China (Nos. 12090040, 12090043, 
11873016, 11873054, 
12303068.B.Z. 
), and the Science Research Grants from the China Manned Space Project 
(NOs. CMS-CSST-2021-A10, 
CMS-CSST-2021-A08). 
XC acknowledges the National Science Fund for Distinguished Young Scholars (No. 12125303 ).
CJ acknowledges the Strategic Pioneer Program on Space Science, Chinese Academy of Sciences (CASs) through grant NOs. XDA15052100, and the Strategic Priority Research Program of the Chinese Academy of Sciences, Grant No. XDB0550200.

Guoshoujing Telescope (the Large Sky Area Multi-Object Fiber Spectroscopic Telescope LAMOST) is a National Major Scientific Project built by CASs. Funding for the project has been provided by the National Development and Reform Commission. LAMOST is operated and managed by the National Astronomical Observatories, CASs. The LAMOST FELLOWSHIP is supported by Special Funding for Advanced Users, budgeted and admin-istrated by Center for Astronomical Mega-Science, CASs. This work is supported by Cultivation Project for LAMOST Scientific Payoff and Research Achievement of CAMS-CAS and Special research assistant program of CASs. 

This paper includes data collected by the TESS mission. Funding for the TESS mission is provided by the NASA Explorer Program. This work presents results from the European Space Agency (ESA) space mission Gaia. Gaia data are being processed by the Gaia Data Processing and Analysis Consortium (DPAC). Funding for the DPAC is provided by national institutions, in particular the institutions participating in the Gaia Multi Lateral Agreement (MLA). The Gaia mission website is \url{https://www.cosmos.esa.int/gaia}. The Gaia archive website is \url{https://archives.esac.esa.int/gaia}. This research made use of the data from the Milky Way Imaging Scroll Painting (MWISP) project, which is a multi-line survey in 12CO/13CO/C18O along the northern galactic plane with PMO-13.7m telescope. The authors thank all the members of the MWISP working group, particularly the staff members at the PMO-13.7m telescope, for their long-term support. MWISP was sponsored by National Key R\&D Program of China with grant 2017YFA0402701 and CAS Key Research Program of Frontier Sciences with grant QYZDJ-SSW-SLH047.

{\it Software or package}: \href{https://github.com/pmaxted/ellc}{ellc \citep{Maxted2016}},
\href{https://github.com/hypergravity/bfosc}{bfosc},
\href{https://github.com/lidihei/pyrafspec}{pyrafspec},
\href{https://laspec.readthedocs.io/en/latest/}{laspec \citep{Zhangbo2021ApJS}},
\href{https://docs.pymc.io/en/v3/index.html}{PyMC3 \citep{salvatier2016}}, 
\href{https://emcee.readthedocs.io/en/stable/}{emcee \citep{emcee2013ascl.soft03002F}},
\href{https://www.astropy.org/}{Astropy \citep{astropy:2013, astropy:2018}},
\href{https://docs.lightkurve.org/}{Lightkurve \citep{lightkurve2018ascl.soft12013L}},
\href{https://dustmaps.readthedocs.io/en/latest/}{dustmaps \citep{Green2018JOSS}},
\href{http://www.star.bris.ac.uk/~mbt/topcat/}{TOPCAT \citep{TOPCAT2005ASPC}}, 
\href{https://iraf-community.github.io/install.html}{iraf \citep{iraf1986SPIE..627..733T, iraf1993ASPC...52..173T}},
\href{https://github.com/python/}{Python}, 
Numpy \citep{Numpy2011CSE....13b..22V, Numpy2020Natur.585..357H}, Scipy \citep{SciPy-NMeth2020, jones_scipy:_2001},
\href{https://pyastronomy.readthedocs.io/en/latest/index.html}{PyAstronomy \citep{pyastronomy}},
\href{https://speedyfit.readthedocs.io/en/stable/index.html}{Speedyfit}.


\appendix
\setcounter{table}{0}
\setcounter{figure}{0}
\renewcommand\thefigure{\thesection.\arabic{figure}}
\renewcommand{\thetable}{A\arabic{table}}
\section{Appendix information}

\subsection{Lomb-Scargle periodogram}\label{sec:periodogram}
\begin{figure}
    \centering
    \plotone{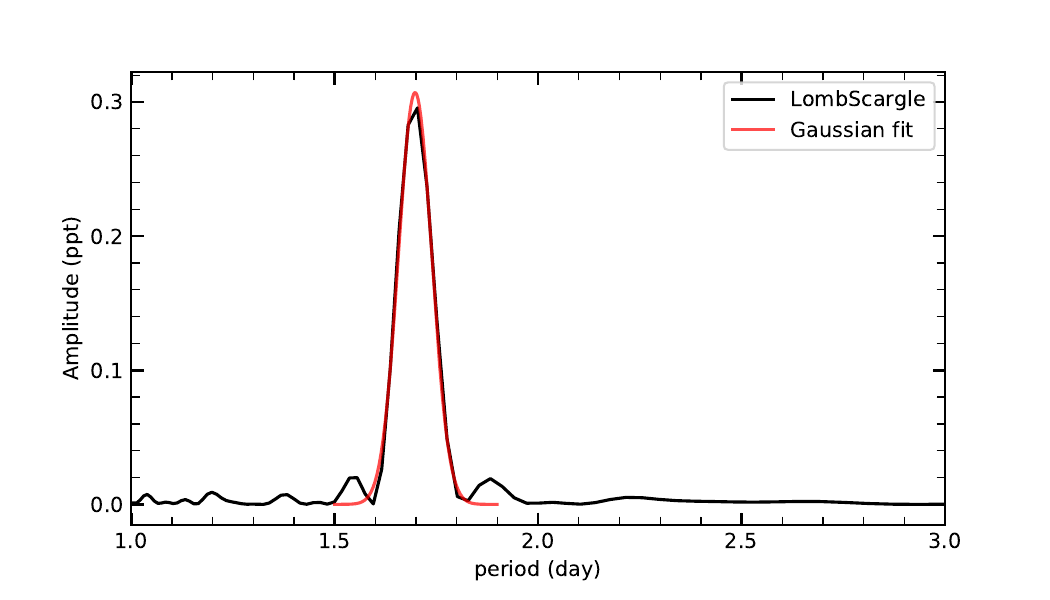}
    \caption{Periodogram of the TESS light curve. The black and red lines are the Lomb-Scargle periodogram and the fitted Gaussian curve, respectively.}
    \label{fig:period}
\end{figure}

A Lomb-Scargle periodogram of the TESS LC shows a peak. So we use the Gaussian function to fit the peak and obtain the peak period is 1.69759(59) day (see Figure~\ref{fig:period}). The peak period and its error are the mean and square root of the estimated covariance of the mean of the Gaussian function. This is performed by a Python function \texttt{curve\_fit}.

\subsection{Observation and data reduction} \label{sec:foscspec}

TYC 3340-2437-1 spectra were collected from three observation facilities, including LAMOST, the Xinglong 2.16-m telescope at the Xinglong station of the National Astronomical Observatories, CAS, and the Lijiang 2.4-m telescope at the Lijiang station of the Yunnan Observatories, CAS. We collected spectra observed in both the low ($R\sim1800$) and medium ($R\sim7500$) resolution surveys from the LAMOST. Additional spectra were observed using the BFOSC E9$+$G10 instrument with a 1.6$''$  on the Xinglong 2.16-m telescope ($R\sim2000$), and the YFOSC E9$+$G10 instrument with a 0.58$''$ slit on the Lijiang 2.4-m telescope with $R\sim2500$. The BFOSC spectra were reduced and extracted by using the developed Python package \texttt{BFOSC}. 
The YFOSC spectra were reduced using \texttt{iraf}. Standard reduction procedures, such as bias subtraction, flat fielding correction, spectral order extraction, and wavelength calibration using the Fe/Ar lamp for BFOSC observations (He/Ne lamp for YFOSC) were included to reduce the data. Spectral orders were combined using the Python package \texttt{pyrafspec}. Flux measurements contaminated by skylight were excluded beyond the wavelength range larger than 7200 \AA.
In Table~\ref{tab:telescope}, we list the telescope, instrument, adopted slit, spectral resolution, selected calibration lamp, wavelength coverage, the number of observations, and the location of the telescope.

\begin{table}
\centering
\begin{tabular}{c|ccccccc} 
\hline\hline
Telescope & Instrument & Slit & $R$&Lamp & Wavelength & $N_{\rm obs}$ & Location\\
          &            & ($''$)&    &    & (\Angstrom) &               \\
\hline
LAMOST    & LRS        &  -   & $\sim1800$ &HgCdNe  &  [3700, 9000] &     1 & Xinglong (China)\\
LAMOST    & MRS        &  -   & $\sim7500$ &ThAr   &  [4900, 5350] and [6300, 6800] &     21& Xinglong (China)\\
The Xinglong 2.16-m & BFOSC/E9+G10 & 1.6 & $\sim2000$& FeAr & [3900, 7200] & 16& Xinglong (China)\\
The Lijiang 2.4-m & YFOSC/E9+G10 & 0.58 & $\sim2500$&HeNe & [5200, 7200] & 2 & Lijiang (China)\\
\hline\hline
\end{tabular}
\caption{Basic information of each observation facility. $R$ is the spectrum resolution. $N_{\rm obs}$ is observation number. Wavelength stands for the effective wavelength coverage. {\it Note}: the effective wavelength coverage of YFOSC spectrum is less than that of BFOSC because the absence of emission lines of HeNe lamp in the wavelength range 3900-5200 \Angstrom. } \label{tab:telescope}
\end{table}

\begin{table}\label{tab:rv}
\centering
\begin{tabular}{cccccccc}
\hline\hline
BJD        & phase   & RV$_{1}$ & RV$_2$  &  RV$_{\rm NaD1}$ & RV$_{\rm NaD2}$ & Telescope/Instrument & Exposure\\
(day)   &         &(${\rm kms^{-1}}$)  &  (${\rm kms^{-1}}$) &  (${\rm kms^{-1}}$)  &  (${\rm kms^{-1}}$)        &        & (s)  \\
\hline

	 2457326.255659 & -0.015005 & $-8.1\pm1.3$ & -- &$-15.1\pm3.2$ & $-9.5\pm3.0$ & LAMOST/LRS & 1800.0 \\
	 2458452.123684 & 0.050773 & $-40.4\pm0.5$ & -- &-- & -- & LAMOST/MRS & 1200.0 \\
	 2458452.139922 & 0.055563 & $-45.5\pm0.7$ & -- &-- & -- & LAMOST/MRS & 1200.0 \\
	 2458452.166600 & 0.063431 & $-52.2\pm1.0$ & $87.0\pm4.9$ &-- & -- & LAMOST/MRS & 1200.0 \\
	 2458452.183383 & 0.068381 & $-57.8\pm0.9$ & $80.3\pm2.8$ &-- & -- & LAMOST/MRS & 1200.0 \\
	 2458452.200883 & 0.073543 & $-64.7\pm0.7$ & $85.5\pm2.3$ &-- & -- & LAMOST/MRS & 1200.0 \\
	 2458452.217098 & 0.078325 & $-64.3\pm0.7$ & $88.5\pm2.6$ &-- & -- & LAMOST/MRS & 1200.0 \\
	 2458452.233534 & 0.083173 & $-68.7\pm0.7$ & $88.0\pm2.6$ &-- & -- & LAMOST/MRS & 1200.0 \\
	 2458482.093813 & -0.109777 & $53.5\pm0.5$ & $-136.4\pm2.8$ &-- & -- & LAMOST/MRS & 1200.0 \\
	 2458482.110016 & -0.104998 & $50.3\pm0.6$ & $-133.5\pm2.9$ &-- & -- & LAMOST/MRS & 1200.0 \\
	 2458482.126219 & -0.100219 & $46.6\pm0.6$ & $-124.9\pm3.2$ &-- & -- & LAMOST/MRS & 1200.0 \\
	 2458482.142410 & -0.095443 & $42.7\pm0.7$ & $-118.7\pm2.6$ &-- & -- & LAMOST/MRS & 1200.0 \\
	 2458482.158613 & -0.090664 & $41.6\pm0.6$ & $-114.1\pm3.0$ &-- & -- & LAMOST/MRS & 1200.0 \\
	 2458508.956791 & -0.186756 & $83.1\pm0.5$ & $-200.5\pm3.0$ &-- & -- & LAMOST/MRS & 900.0 \\
	 2458508.969521 & -0.183001 & $82.4\pm0.5$ & $-204.2\pm4.4$ &-- & -- & LAMOST/MRS & 900.0 \\
	 2458508.982240 & -0.179250 & $81.9\pm0.5$ & $-193.9\pm2.1$ &-- & -- & LAMOST/MRS & 900.0 \\
	 2458879.951649 & 0.235213 & $-117.9\pm0.5$ & $177.8\pm4.3$ &-- & -- & LAMOST/MRS & 1200.0 \\
	 2458879.967930 & 0.240015 & $-115.7\pm0.6$ & $183.0\pm3.0$ &-- & -- & LAMOST/MRS & 1200.0 \\
	 2458879.984165 & 0.244803 & $-115.4\pm0.6$ & $183.1\pm4.2$ &-- & -- & LAMOST/MRS & 1200.0 \\
	 2459188.120908 & 0.127274 & $-89.0\pm0.8$ & $129.7\pm4.3$ &-- & -- & LAMOST/MRS & 1200.0 \\
	 2459188.137111 & 0.132053 & $-91.4\pm0.8$ & $136.0\pm4.4$ &-- & -- & LAMOST/MRS & 1200.0 \\
	 2459188.153317 & 0.136833 & $-93.0\pm0.8$ & $140.3\pm5.9$ &-- & -- & LAMOST/MRS & 1200.0 \\
	 2459200.066757 & -0.349393 & $66.1\pm26.1$ & -- &$-11.8\pm3.6$ & $-16.6\pm3.8$ & Xinglong 2.16m/BFOSC & 900.0 \\
	 2459200.206025 & -0.308317 & $74.9\pm5.7$ & -- &$-22.7\pm0.8$ & $-21.7\pm0.9$ & Xinglong 2.16m/BFOSC & 1200.0 \\
	 2459200.376090 & -0.258158 & $83.0\pm27.1$ & -- &$-6.5\pm2.0$ & $-8.5\pm2.0$ & Xinglong 2.16m/BFOSC & 900.0 \\
	 2459201.057383 & -0.057216 & $-17.0\pm5.9$ & -- &$-45.5\pm0.7$ & $-43.8\pm0.7$ & Xinglong 2.16m/BFOSC & 1200.0 \\
	 2459201.190748 & -0.017881 & $-24.3\pm7.2$ & -- &$-35.3\pm1.1$ & $-33.4\pm1.1$ & Xinglong 2.16m/BFOSC & 1200.0 \\
	 2459201.353614 & 0.030155 & $-43.2\pm18.7$ & -- &$-14.8\pm3.4$ & $-14.2\pm2.5$ & Xinglong 2.16m/BFOSC & 1200.0 \\
	 2459202.048319 & 0.235053 & $-120.8\pm7.0$ & -- &$-23.8\pm0.9$ & $-18.7\pm0.9$ & Xinglong 2.16m/BFOSC & 1200.0 \\
	 2459202.175862 & 0.272670 & $-113.5\pm13.2$ & -- &$-14.1\pm1.3$ & $-15.9\pm1.3$ & Xinglong 2.16m/BFOSC & 1200.0 \\
	 2459202.369295 & 0.329722 & $-89.2\pm14.2$ & -- &$-22.3\pm1.4$ & $-20.0\pm1.3$ & Xinglong 2.16m/BFOSC & 1200.0 \\
	 2459203.057032 & -0.467436 & $-10.0\pm4.8$ & -- &$-27.9\pm0.8$ & $-30.4\pm0.8$ & Xinglong 2.16m/BFOSC & 1200.0 \\
	 2459203.179065 & -0.431443 & $15.3\pm6.7$ & -- &$-30.4\pm0.8$ & $-29.6\pm0.8$ & Xinglong 2.16m/BFOSC & 1200.0 \\
	 2459203.368586 & -0.375545 & $48.6\pm14.2$ & -- &$-19.6\pm1.4$ & $-17.0\pm1.3$ & Xinglong 2.16m/BFOSC & 1200.0 \\
	 2459219.132717 & 0.273959 & $-127.4\pm2.6$ & -- &$-18.4\pm0.8$ & $-17.5\pm0.9$ & Lijiang 2.4m/YFOSC & 900.0 \\
	 2459219.942077 & -0.487326 & $4.4\pm13.6$ & -- &$3.7\pm1.4$ & $-0.5\pm1.4$ & Xinglong 2.16m/BFOSC & 1200.0 \\
	 2459219.999129 & -0.470499 & $4.4\pm4.3$ & -- &$-18.2\pm1.3$ & $-19.5\pm1.2$ & Lijiang 2.4m/YFOSC & 1200.0 \\
	 2459220.111641 & -0.437315 & $0.2\pm15.6$ & -- &$-34.3\pm2.0$ & $-37.2\pm2.1$ & Xinglong 2.16m/BFOSC & 1200.0 \\
	 2459530.227730 & 0.028948 & $-32.1\pm0.8$ & -- &-- & -- & LAMOST/MRS & 900.0 \\
	 2459530.239472 & 0.032411 & $-33.4\pm0.9$ & -- &-- & -- & LAMOST/MRS & 900.0 \\
	 2459530.251216 & 0.035875 & $-35.5\pm0.9$ & -- &-- & -- & LAMOST/MRS & 900.0 \\
	 2459590.021107 & -0.335474 & $77.2\pm0.7$ & $-200.5\pm4.0$ &-- & -- & LAMOST/MRS & 1200.0 \\
	 2459590.042371 & -0.329202 & $78.7\pm0.7$ & $-193.1\pm3.3$ &-- & -- & LAMOST/MRS & 1200.0 \\
	 2459590.057588 & -0.324714 & $80.9\pm0.7$ & $-210.8\pm7.3$ &-- & -- & LAMOST/MRS & 1200.0 \\
	 2459605.016181 & 0.087204 & $-69.0\pm0.7$ & $100.9\pm3.1$ &-- & -- & LAMOST/MRS & 1200.0 \\
	 2459605.031397 & 0.091692 & $-73.4\pm0.8$ & $99.4\pm3.1$ &-- & -- & LAMOST/MRS & 1200.0 \\
	 2459605.046618 & 0.096181 & $-74.8\pm0.7$ & $113.0\pm4.1$ &-- & -- & LAMOST/MRS & 1200.0 \\
	 2459619.014359 & 0.215854 & $-113.5\pm0.5$ & $185.3\pm3.2$ &-- & -- & LAMOST/MRS & 1200.0 \\
	 2459619.029577 & 0.220343 & $-117.0\pm0.5$ & $173.3\pm3.7$ &-- & -- & LAMOST/MRS & 1200.0 \\
	 2459619.044791 & 0.224830 & $-116.9\pm0.5$ & $178.7\pm3.8$ &-- & -- & LAMOST/MRS & 1200.0 \\
\hline\hline
\end{tabular}
\caption{Radial velocities, RV$_1$ is radial velocity of stat B$_\BOa$, which is measured by CCF. RV$_2$ is the radial velocity of star B$_\BOb$, which is measured by MCMC approach (details see Section~\ref{sec:fit_rv2}). RV$_{\rm NaD1}$ and RV$_{\rm NaD2}$ are RVs of ISM absorption lines Na D1 (5890 \Angstrom) and Na D2 (5896 \Angstrom), which are measured by the Gaussian fit. 
}
\end{table}

\subsection{Atmosphere Parameters of Star B$_\BOa$}\label{sec:BOa_atm}

Despite the spectral resolution of LAMOST-LRS being $R \sim 1800$, for early-type stars, the uncertainties of atmosphere parameters obtained by LAMOST-LRS spectra are better than that of LAMOST-MRS spectra ($R\sim 7500$) since the distinctness optical feature of early-star concentrate on the wavelength range of [3700, 5000] $\Angstrom$ (the wavelength coverage shown in Table~\ref{tab:telescope}). The wavelength of LAMOST-LRS spectrum range from [3700, 9000] $\Angstrom$, while LAMOST-MRS wavelength only is in the intervals of [4900, 5350] $\Angstrom$ (blue arm) and [6300, 6800] $\Angstrom$ (red arm). 
Moreover, the signal-to-noise ratio (SNR) of BFOSC spectra are less than the SNR of LAMOST-LRS spectrum. Therefore, we used LAMOST-LRS spectrum, which has been corrected for its wavelength to the rest frame, to measure the atmosphere parameters.  We perform a Markov chain Monte Carlo approach (MCMC) to determine the parameters: effective temperature $T_{\rm eff}$, surface gravity $\log g$, metallicity $Z$, and projected rotation velocity \vsini. The posterior distribution is,
\begin{equation}
\begin{split}
    &ln [\mathcal{P}(T_{\rm eff}, \log g, Z, v\sin i \mid f, f^{\rm syn} ) ]\\
    & \propto ln [\mathcal{P}(f, f^{\rm syn} \mid T_{\rm eff}, \log g, Z, v\sin i ) ]\\
    &=-\frac{1}{2}\sum_{i=1}\frac{[f_{i}-f_{i}^{\rm syn}]^2}{\sigma^2_{f_{i}}} -\frac{1}{2}\sum_{n=1}\ln(2\pi\sigma_{f_i}^2),
\end{split}
\end{equation}
where $f_i$ and $\sigma_{f_i}$ represent flux and flux error of each pixel of LAMOST-LRS spectrum, respectively. Pixels within the wavelength range of ($3760 < \lambda < 3900$) or ($4000 < \lambda < 4400$) or ($4460 < \lambda < 5000$) were used to estimate the atmosphere parameters (as the residual spectrum shown in the bottom panel of Figure~\ref{fig:spec_lrs}). $f^{\rm syn}$ denote the synthetic spectrum interpolated using the Stellar Label Machine (SLAM) \citep{Zhangbo2020ApJS, Guoyanjun2021ApJS} with a TLUSTY spectra grid.

SLAM is a data-driven method based on support vector regression, which can be used to interpolate spectra with a trained model. The model was trained by spectra of a subgrid of TLUSTY OSTAR2002. We have selected a subset: $27500 \leq T_{\rm eff} \leq 37500$ K with a step of 2500, $3.25 \leq \log g \leq 4.5$ dex with a step of 0.25, four chemical compositions of 0.2, 0.5, 1 and 2 $Z_\odot$, and microturbulence velocity $\xi_{\rm t} = 10$ \kms, from the TLUSTY OSTAR2002 grid\citep{Lanz2003ApJS}. Each spectrum within the subset grid was broadened with projected rotational velocities $0\leq v\sin i \leq 200$ \kms with a step of 10 (using the broadening kernel described by Eq. 18.14 of \cite{gray_2021} ). The limb-darkening coefficient was fixed $\epsilon = 0.2$. Subsequently, the broadened spectra were degraded to an instrument resolution of 1800 and then resampled with a wavelength range of $3700 \leq \lambda <6700$ $\Angstrom$, with logarithmic steps of 0.0001 based on a base 10 logarithm. This corresponds to the dispersion per pixel of LAMOST-LRS spectra.

\begin{figure}[htbp]
    \centering
        \includegraphics[width=0.8\textwidth]{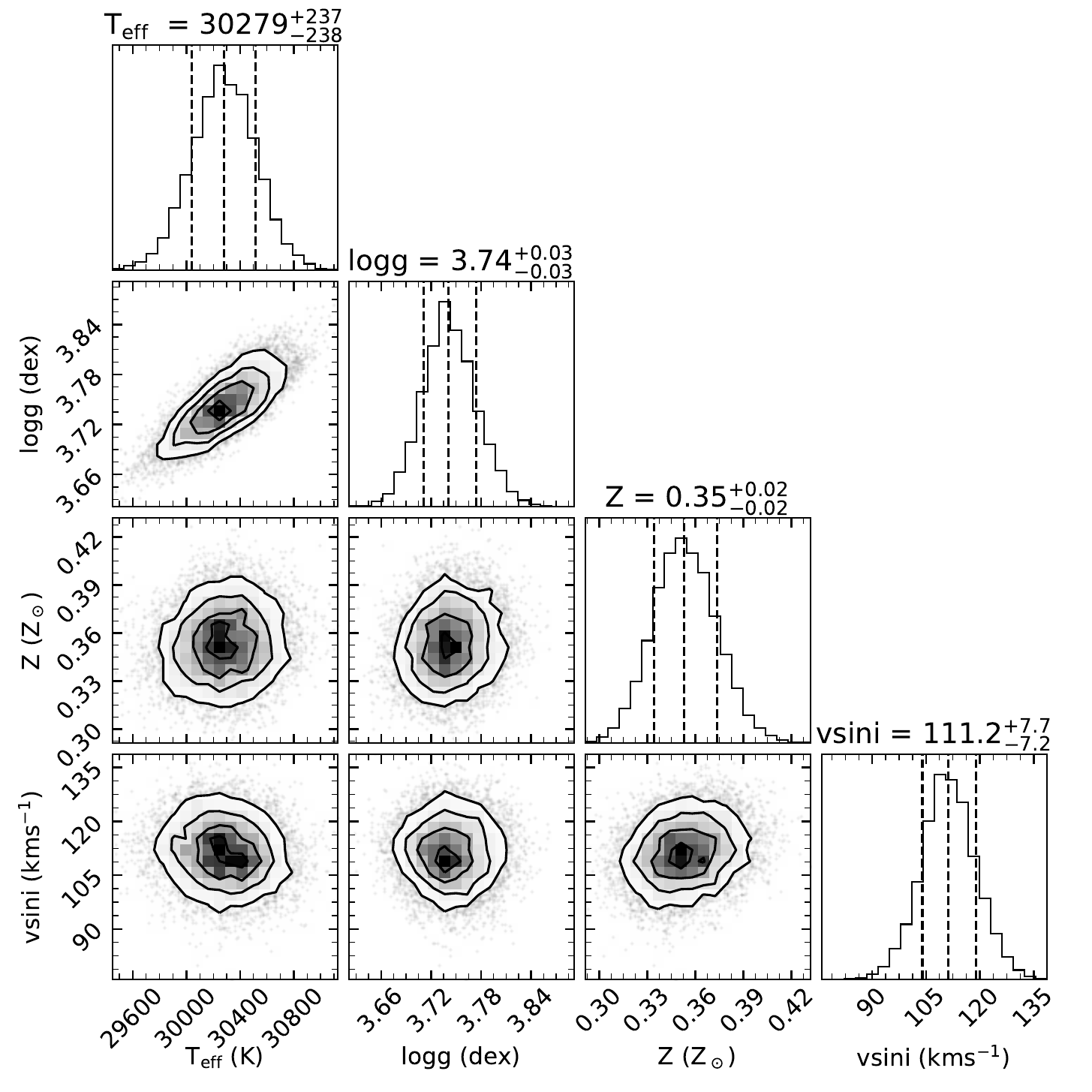}
    \caption{ Corner plot for LAMOST-LRS spectrum MCMC fit. The posterior probability distribution for the parameters derived from the MCMC fit to LAMOST-LRS spectrum. }
    \label{fig:lamost-lrs_a}
\end{figure}

We obtain $T_{\rm eff} = 30279\pm 230$ K, $\log g = 3.74 \pm 0.03$ dex, $Z = 0.35\pm 0.02~Z_\odot$, and $v\sin i = 111 \pm 8$ ${\rm kms^{-1}}$. Their errors are relatively small (see Figure~\ref{fig:lamost-lrs_a}). To assess this, we applied the Cram\'er-Rao bound \citep{Cramer1946, Rao1945} for the trained SLAM model. The theoretically achievable uncertainties of $T_{\rm eff}$, $\log g$, $Z$ and $v\sin i$ were calculated to be 210 K, 0.027 dex, 0.019 $Z_{\odot}$ and 6.5 \kms, respectively. They are comparable to the errors of the MCMC approach. Therefore, the low errors are reasonable, since only precision measurements were taken into account.

\subsubsection{Contamination of the other stars}\label{sec:spec_contaminaion}
The spectrum of star $B_\BOa$ is contaminated by the other three stars.
To simulate the LAMOST-LRS spectrum of the quadruple system, we assigned the following parameters: for star $B_\BOa$, $T_{\rm eff} = 30000$ K, $\log g = 3.7$ dex, $Z = 0.35$ $Z_\odot$, $v\sin i = 0$ \kms, and a radius of $7.8$ $R_\odot$; for the other three stars, $T_{\rm eff} = 19000$ K, $\log g = 4.3$ dex, $Z = 0.35$ $Z_\odot$, $v\sin i = 0$ \kms, and a radius of $3$ $R_\odot$. The radial velocities of stars $B_\BOa$ and $B_\BOb$ were set to 9 and -18 \kms, respectively. The semi-amplitudes of the radial velocities of stars $B_\BBa$ and $B_\BBb$ were 83 and 215 \kms, respectively.

We varied the center mass radial velocity of binary $B_\BB$ relative to binary $B_\BO$ from -20 to 20 with a step of 10 \kms and the orbital phase of binary $B_\BB$ from 0 to 0.9 with a step of 0.1. For each center mass and orbital phase point of binary $B_\BB$, we generated 100 spectra and randomly resampled flux to ensure that each pixel of the mock spectra had the same SNR as the LAMOST-LRS spectrum using a Gaussian distribution. Subsequently, the atmospheric parameters of the mock spectra were predicted using SLAM.

\begin{figure}[htbp]
    \centering
        \includegraphics[width=0.8\textwidth]{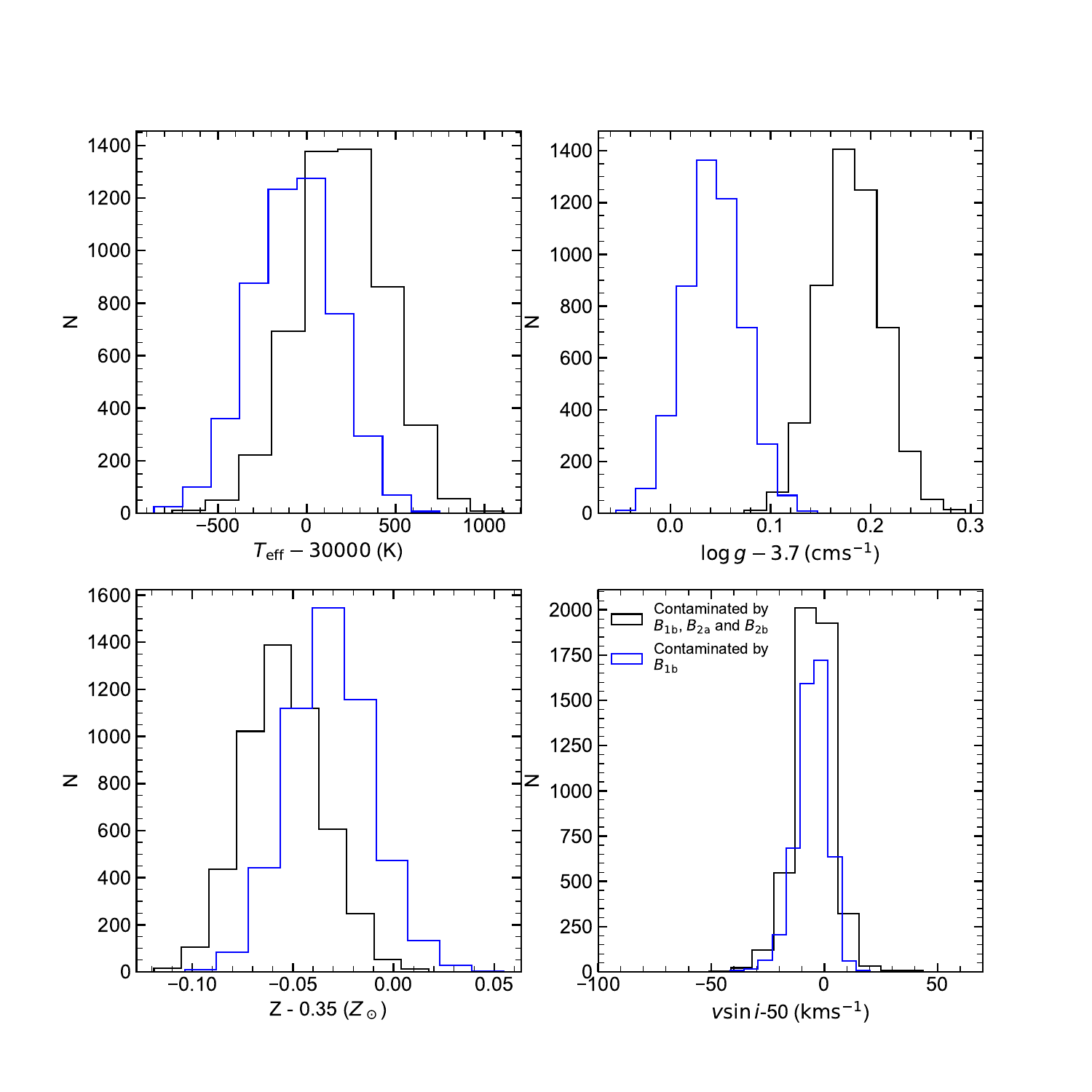}
    \caption{The atmosphere parameters biases of star $B_\BOa$ due to the contamination of spectrum by the components stars. The blue and black lines represent the bias caused by only star $B_\BOb$ and the other three stars, respectively.}
    \label{fig:spec_offset}
\end{figure}

We observed biases of about 200 K, 0.18 dex, -0.05 $Z_{\odot}$, -5 \kms in $T_{\rm eff}$, $\log g$, $Z$ and $v\sin i$ respectively, for star $B_\BOa$ (see back line of Figure~\ref{fig:spec_offset}). When considering only the contamination of the spectrum by star $B_\BOa$, the biases in $T_{\rm eff}$, $\log g$, $Z$ and $v\sin i$ were about -72 K, 0.04 dex, -0.03 $Z_{\odot}$ and -5 \kms (see blue line of Figure~\ref{fig:spec_offset}). We also observed that the biases are almost invariable for all orbital phases and center mass velocities of binary $B_\BB$.

For a more comprehensive evaluation that considers both precision and accuracy, we have accepted the uncertainties reported by \cite{Guoyanjun2021ApJS} for $\sigma(T_{\rm eff}) = 1600$ K, $\sigma(\log g) = 0.25$ dex, and $\sigma(v\sin i) = 42$ \kms. It should be noted that the resulting estimated metallicity could potentially be unrealistic due to the weakness of metallicity lines in the optical wavelength range, especially given the high temperature of the spectrum.

\begin{figure}[htbp]
    \centering
        \includegraphics[width=1\textwidth]{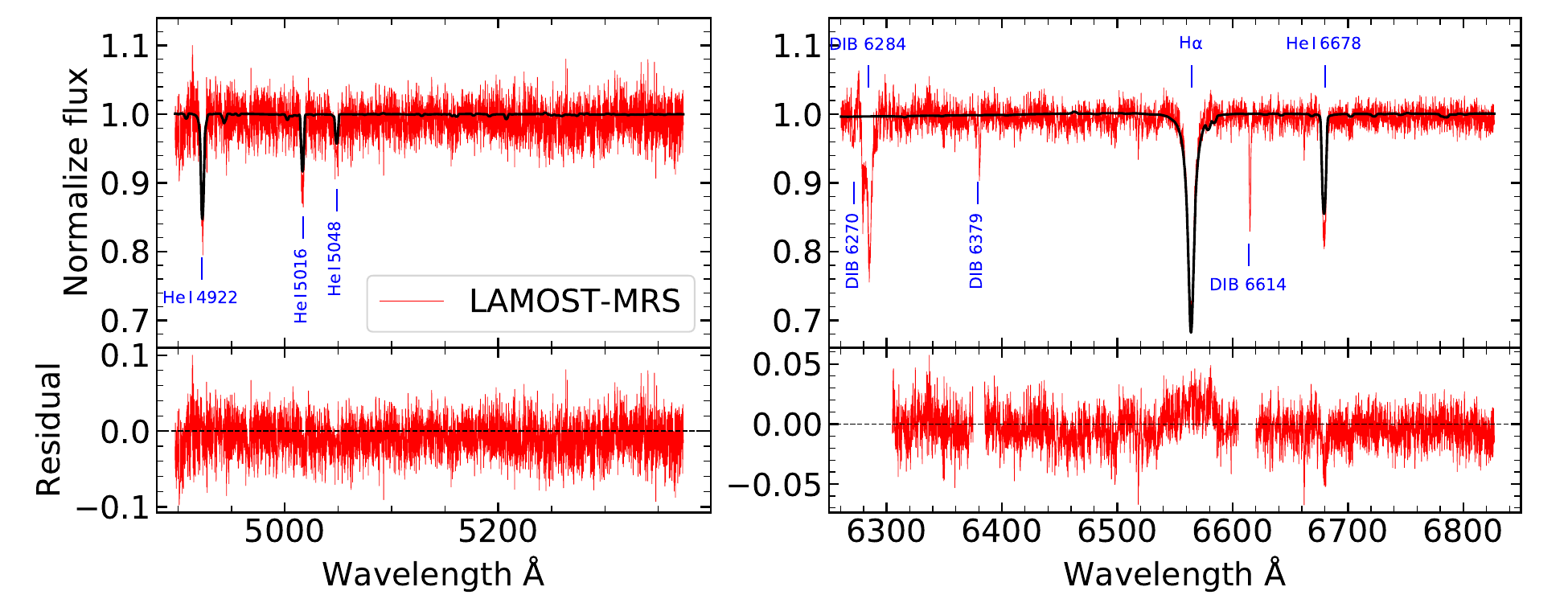}
    \caption{The LAMOST-MRS spectrum. The spectrum was observed at the orbital phase of 0.028665 of binary B$_\BO$. The right and left panels are blue and red arm spectra of LAMOST-MRS. The black line is interpolated by using the best-fitting atmosphere parameters of the LAMOST-LRS spectrum. The left bottom panel contains blank wavelengths where the DIBs are located.}
    \label{fig:lamost-mrs}
\end{figure}

\subsection{Vicinity Eclipse Binary?}\label{sec:vicinity_eclipse}
Assuming that binary B$_\BB$ is a vicinity eclipse binary with an inclination angle of $90$\degree, the primary eclipse relative depth of B$_\BB$ is
\begin{equation}\label{delta1}
\begin{split}
    \Delta_1 &= 1 - \frac{L_\BO + L_\BBb}{L_\BO + L_\BBa + L_\BBb}\\
             & = \frac{L_\BBa}{L_\BO + L_\BBa + L_\BBb},
\end{split}
\end{equation}
where $L_\BO$, $L_\BBa$, and $L_\BBb$ are luminosity of binary B$_\BO$, star B$_\BBa$, and star B$_\BBb$, respectively. Similarly, the secondary relative eclipse depth is
\begin{equation}\label{delta2}
    \Delta_2 = \frac{L_\BBb}{L_\BO + L_\BBa + L_\BBb}.
\end{equation}
Combining Equation (\ref{delta1}) and (\ref{delta2}), we can figure out that the luminosity of binary B$_\BB$ is
\begin{equation}\label{eq:LBB}
\begin{split}
    L_\BB &= L_\BBa + L_\BBb \\
          &= (\Delta_1 + \Delta_2)(L_\BO + L_\BB)\\
          &>  (\Delta_1 + \Delta_2)L_\BO,
\end{split}
\end{equation}
so that the magnitude of binary B$_\BB$ is
\begin{equation}\label{eq:magBB}
\begin{split}
    m_\BB &< -2.5\log (\Delta_1 + \Delta_2) + m_\BO\\
          & = -2.5\log (\Delta) + m_\BO.
\end{split}
\end{equation}

\begin{figure}
\gridline{\fig{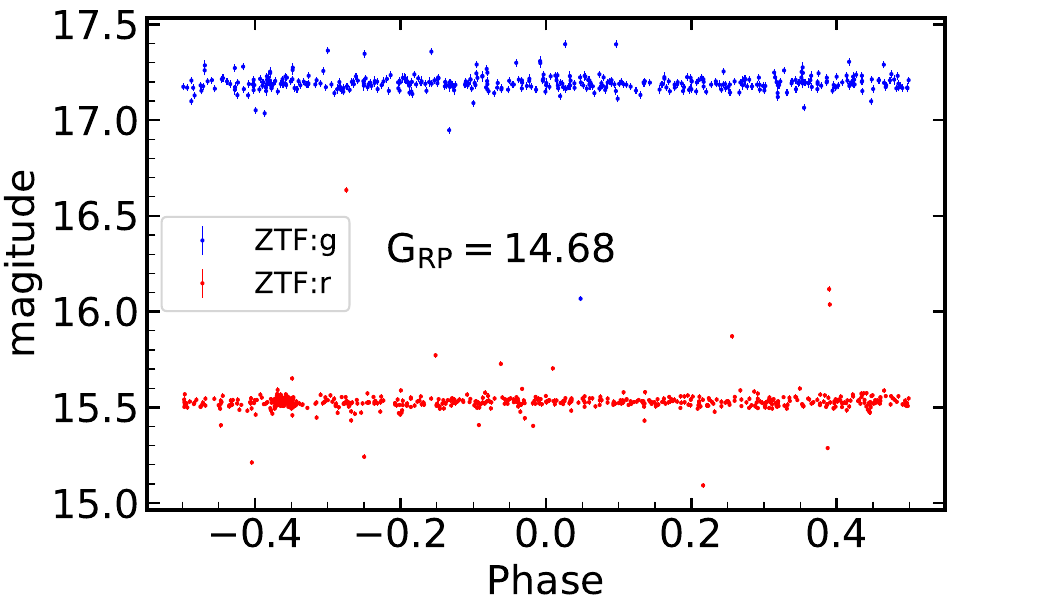}{0.4\textwidth}{(a)}
          \fig{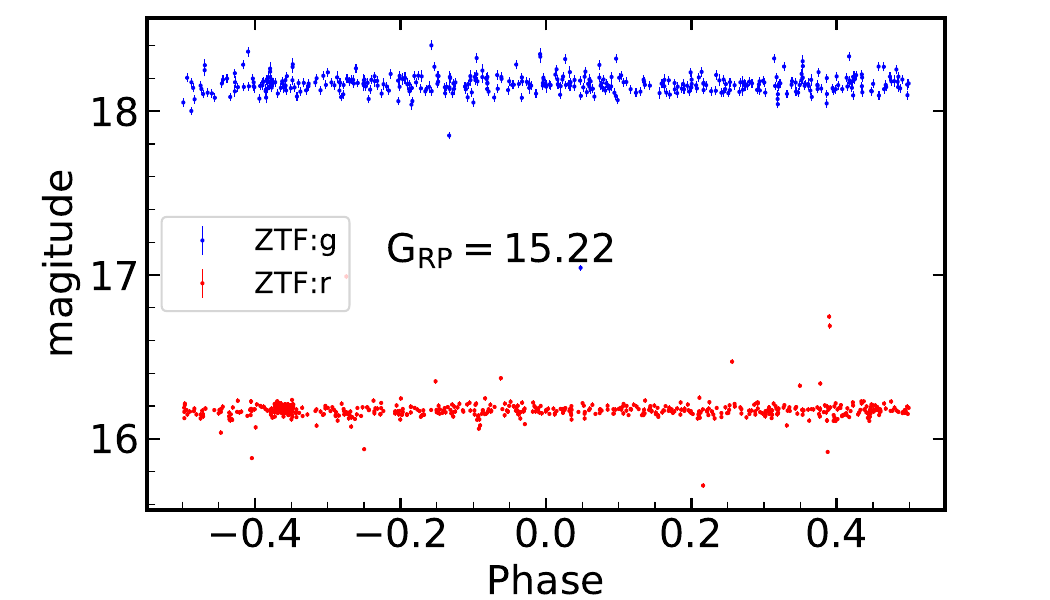}{0.4\textwidth}{(b)}
          }
\caption{Phase-folded ZTF LCs of the two sources with $G_{RP} = 14.68$ ($\alpha=62.7854177\degree, \delta=50.6993595\degree$) and $G_{RP} = 15.22$ ($\alpha=62.7887683\degree, \delta=50.6935972\degree$). The folded period is 2.4378 days. The blue and red dots are the magnitude of $g$ and $r$ bands of ZTF LC, respectively. 
\label{fig:ztflc}}
\end{figure}

In the top panel of Figure~\ref{fig:tesslc} (c), we can see that $\Delta \in (0.01, 0.02)$. Taking into account the completeness of the observation and the response function of the TESS filter, we use {\it Gaia} $G\RP$ as a reference. The faintest available magnitude of binary B$_\BB$ is 15.12 mag based on the $G\RP = 10.12$ mag of \TYC. Crossing match {\it Gaia} DR3 with an angular radius of 60$''$ at the center of \TYC{}, we find most of the sources are fainter than 17 mag, expect two sources have magnitude $G\RP = 14.68$ ($38''$ away; $\alpha=62.7854177\degree, \delta=50.6993595\degree$) and $G\RP = 15.22$ ($54''$ away; $\alpha=62.7887683\degree, \delta=50.6935972\degree$). We obtain their light curves from ZTF DR9 and then fold them into the period of 2.43781 days (see Figure~\ref{fig:ztflc}). The folded light curves are almost uniformly distributed in the period phase, indicating that they do not have the orbital period of 2.43781 days. Binary B$_\BB$, therefore, is impossible to be a vicinity eclipse binary. 


\subsection{Posterior distribution of fitting light curve and radial velocity curve}\label{sec:mcmclcrv}

In the process of fitting, we fixed the orbital periods of binary B$_\BO$ and B$_\BB$, the eccentricity and transit time of B$_\BB$, and the effective temperature of star B$_\BOa$ ($P_\BO = 3.390497$ days, $P_\BB = 2.43781$ days, $e_\BB=0$, $T_{\rm c\BB} = 2458817.49056$ BJD, $T_{\rm eff\BOa}=30279$ K). The posterior distribution of MCMC fit by simultaneously using LC and radial velocity curves of binary B$_\BO$.
\begin{equation}\label{eq:lnlk_lcrv}
    \mathcal{P}(\theta \mid l, v_\BOa , v_\BOa) \propto 
    \mathcal{P}(l, v_\BOa , v_\BOa \mid \theta) = 
    \mathcal{L}(l \mid \theta)
    \mathcal{L}(v_\BOa \mid \theta)
    \mathcal{L}(v_\BOb \mid \theta)
    \mathcal{P}(\theta)
\end{equation}
where $\theta = (T_{\rm c\BO}, \sqrt{e_\BO}\cos\omega_\BOa, \sqrt{e_\BO}\sin\omega_\BOa, q_\BO, K_\BO, v_{\gamma\BO}, s, R_\BOa, r_\BO, \log g_\BOa, H_\BOb, i_\BB, L_\BBa, R_\BBa, L_{\BBb}, R_\BBb)$.
$\ T_{\rm c}$ is the time of transit; $e$ is eccentricity; $\omega$ is the longitude of periastron; $q$ is the mass ratio ($q_\BO = m_\BOb m^{-1}_\BOa$); $K$ is radial velocity semi-amplitude; $v_\gamma$ is systemic velocity; $R$ is star radius; $r_\BO$ is radius ratio ($r_\BO = R_\BOb R^{-1}_\BOa$); $g_\BOa$ is the gravity of star B$_\BOa$; $H_\BOb$ represents the albedo effect of star B$_\BOb$, which influences the minima of ellipsoidal light curve. As shown in the middle panel of Figure~\ref{fig:tesslc} (b), the minimum at the phase of 0 is less than the minimum at the phase of 0.5 (or -0.5).
$i_{\BB}$ is orbital inclination of binary B$_\BB$;
$L$ is luminosity; 
$s$ is an additional “jitter” variance to description system errors of radial velocities or radial velocity model (In this case, $v_\BOa$ and $v_\BOb$ might be influenced by the binary B$_\BB$). $\mathcal{L}(x | \theta)$ is the likelihood function of $x$, $x$ are the TESS light curve, radial velocities of star B$_\BOa{}$ and star B$_\BOb{}$.
\begin{equation}\label{eq:likelihoodlc}
  \ln[\mathcal{L}(l \mid \theta)] = 
  -\frac{1}{2}\sum_{n=1}\frac{[l_n-l(t_n; \theta_l)]^2}{\sigma^2_{l_n}} -\frac{1}{2}\sum_{n=1}\ln(2\pi\sigma_{l_n}^2),
\end{equation}
where $\theta_l = (T_{\rm c\BO}, \sqrt{e_\BO}\cos\omega_\BOa, \sqrt{e_\BO}\sin\omega_\BOa, q_\BO, K_\BO, R_\BOa, r_\BO, \log g_\BOa, H_\BOb, i_\BB, L_\BBa, R_\BBa, r_{\rm sb\BB}, R_\BBb)$; $l_n$ is the observed light curve flux at time $t_n$. $\sigma_{l_n}$ is error of $l_n$. $l(t_n; \theta_v)$ is model light curve,
\begin{equation}\label{eq:l}
   l(t_n; \theta_v) = l'_\BO(L^{\rm TESS}_\BOa + L^{\rm TESS}_\BOb) +l'_\BB(L^{\rm TESS}_\BBa + L^{\rm TESS}_\BBb)
\end{equation}
$l'$ is calculated by using Python package \texttt{ellc}; $L^{\rm TESS}$ represents the surface brightness through the TESS filter.
Doppler boosting factors are fixed as 4.96 and 5.01 for star B$_\BOa$ and B$_\BOb$, respectively. Limb darkening coefficients follow the four-term law of \cite{Claret2017}, and are interpolated using table 28. Gravity darkening coefficients are from table 29 of \cite{Claret2017}.

\begin{equation}\label{eq:likelihoodrvk1k2}
  \ln[\mathcal{L}(v_y \mid \theta)] = 
  -\frac{1}{2}\sum_{n=1}\frac{[v_{yn}-v_y(t_n; \theta_v)]^2}{\sigma^2_{v_{yn}}+s^2} -\frac{1}{2}\sum_{n=1}\ln [2\pi(\sigma_{v_{yn}}^2+s^2)],
\end{equation}
where $y$ = \BOa~or \BOb, $v_{yn}$ is the measured radial velocity of star B$_\BOa$ or B$_\BOb$, $\theta_v = (T_{\rm c\BO}, \sqrt{e_\BO}\cos\omega_\BOa, \sqrt{e_\BO}\sin\omega_\BOa, q_\BO, K_\BOa, v_{\gamma\BO}, s)$, $v_y(t_n; \theta_v)$ is the Kepler orbital velocity.

\begin{figure*}
    \centering
    \includegraphics[width=0.95\linewidth]{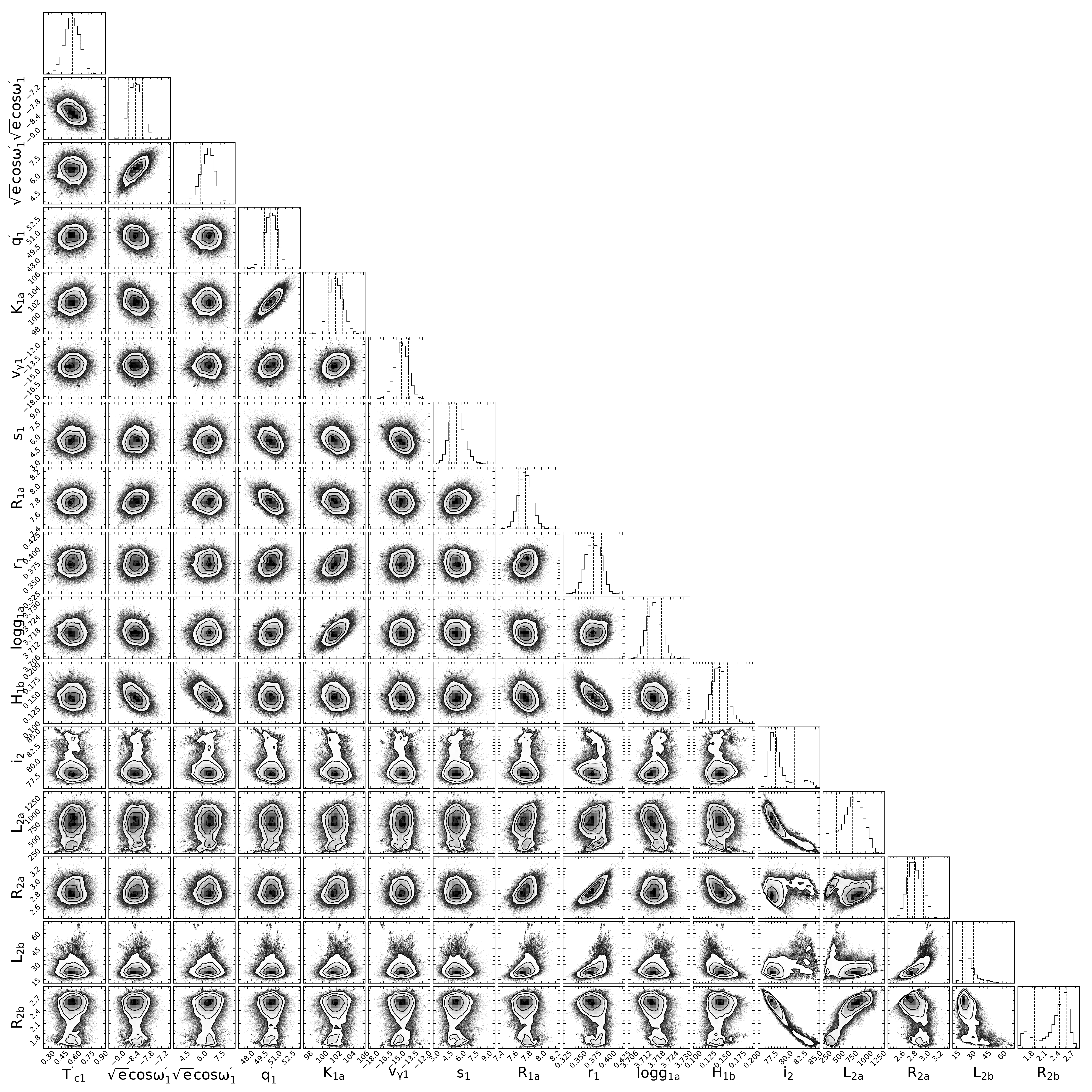}
    \caption{The posterior probability distribution for the parameters derived from the MCMC fit to the TESS light curve and radial velocity curve of binary B$_\BO$. Note: $T'_{\rm c1} = 100(T_{\rm c1} - 2458821.51)$ is in units of BJD; $\sqrt{e}\cos \omega_{1}' = 100\sqrt{e}\cos \omega_{1}$; $\sqrt{e}\cos \omega_{1}' = 100\sqrt{e}\sin \omega_{1}$; $q'_1 = 100q_1$; $K_\BOa$, $v_{\gamma 1}$ and $s_1$ are in units of \kms; $R_\BOa$, $R_\BBa$ and $R_\BBb$ are in units of $R_\odot$; $g_{\BOa}$ in units of cm s$^{-1}$; $L_\BBa$ and $L_\BBb$ are in units of $L_\odot$; $i_2$ is in units of degree; $r_1 = R_{\BOb} R_\BOa^{-1}$; $H_{\BOb}$ is albedo effect of star B$_\BOb$.
    }
    \label{fig:corner_lcrv}
\end{figure*}

In order to effectively constrain MCMC fit, we incorporate prior probability density functions for certain parameters that have been measured through spectra and the star B$_\BOa$ radial velocity curve fitting (see Section~\ref{sec:rv}),
\begin{equation}\label{eq:priorlcrv}
\begin{split}
    \mathcal{P}(\theta) = 
    &\mathcal{P}(T_{\rm c\BO})\mathcal{P}(\sqrt{e_\BO}\cos\omega_\BOa)\mathcal{P}(\sqrt{e_\BO}\sin\omega_\BOa)\mathcal{P}(q_\BO)\mathcal{P}(K_{\BOa})\mathcal{P}(v_{\gamma\BO})\mathcal{P}(s)\\
    &\mathcal{P}(R_\BOa)\mathcal{P}(r_\BO)\mathcal{P}(\log g_\BOa)\mathcal{P}(H_\BOb)\mathcal{P}(i_\BB)\mathcal{P}(L_\BBa)\mathcal{P}(R_\BBa)\mathcal{P}(L_\BBb)\mathcal{P}(R_\BBb)    
\end{split} 
\end{equation}
where
\begin{equation}
    \mathcal{P}(T_{\rm c\BO}) = \mathcal{U}(1821.45, 1821.55)\ ({\rm + 2457000\ day}),
\end{equation}
\begin{equation}
    \mathcal{P}(\sqrt{e_\BO}\cos\omega_\BOa) = \mathcal{U}(-0.32, 0.32),
\end{equation}
\begin{equation}
    \mathcal{P}(\sqrt{e_\BO}\sin\omega_\BOa) = \mathcal{U}(-0.32, 0.32),
\end{equation}
\begin{equation}
    \mathcal{P}(q_\BO) = \mathcal{U}(0.2, 0.7),
\end{equation}
\begin{equation}
    \mathcal{P}(K_\BOa) = \mathcal{U}(80, 140)\ ({\rm km\,s^{-1}}),
\end{equation}
\begin{equation}
    \mathcal{P}(v_{\gamma \BO}) = \mathcal{U}(-30, 10)\ ({\rm km\,s^{-1}}),
\end{equation}
\begin{equation}
    \mathcal{P}(s) = \mathcal{U}(0, 50)\ ({\rm km\,s^{-1}}),
\end{equation}
\begin{equation}
    \mathcal{P}(R_\BOa) = \mathcal{U}(5, 12)\ (R_\odot),
\end{equation}
assuming that \thestar{} is a single star of B$_\BOa$, we obtain a radius of about 10 $R_\odot$ by fitting the spectra energy distribution and combining its {\it Gaia} parallax. Assuming that its four components have the same effective temperature and radius, we can obtain a lower limit of 5 $R_\odot$ for star B$_\BOa$.
\begin{equation}
    \mathcal{P}(r_\BO) = \mathcal{N}(r'_\BO, 0.12);\ r\in [0.1, 1],
\end{equation}
assuming that star B$_\BOb$ is a main sequence star, we can interpolate its luminosity from isochrones of PARSEC for a given mass. Provided with $R_\BOa$, $\log g_\BOa$ and $q_\BO$, it becomes feasible to compute the mass $m_\BOb$ and then the luminosity $L_\BOb$ of star B$_\BOb$. Subsequently, $T_{\rm eff\BOb}$ and $\log g_\BOb$ can be calculated with $L_\BOb$, $m_{\BOb}$, $R_\BOa$ and radius ratio $r_\BO$. Using these values, $r'_\BO$ can be interpolated from the radius ratio grid (Table~\ref{tab:R2tR1}) that varies with $T_{\rm eff\BOb}$ and $\log g_\BOb$ (see Section~\ref{sec:fit_rv2}). This provided an additional condition to constrain radius ratio $r_\BO$. To be conservative, we choose the maximum error of 0.12. 
\begin{equation}
\mathcal{P}(\log g_\BOa) = \mathcal{N}(3.74, 0.25);\ \log g_\BOa\in [3.5, 4.2]\ ({\rm dex}),
\end{equation}
\begin{equation}
\mathcal{P}(H_\BOb) = \mathcal{U}(0, 1),
\end{equation}
from the middle panel of Figure~\ref{fig:tesslc} (b), we found that the flux at the phase of 0 is less than the flux at the phase of 0.5 (-0.5). It indicates that the hemisphere of star B$_\BOb$ towards star B$_\BOa$ is irradiated by the brighter star B$_\BOa$, therefore, we set its albedo effect parameter to vary from 0 to 1.
\begin{equation}
    \mathcal{P}(i_\BB) = \mathcal{U}(30, 90)\ (\degree),
\end{equation}
\[
    \mathcal{P}(L_\BBa)= 
\begin{cases}
    1,                    & \text{if } L_\BBa < L_\BOb,  \\
    0,              & \text{otherwise}.
\end{cases}
\]
\[
    \mathcal{P}(R_\BBa)= 
\begin{cases}
    1,                    & \text{if } R_\BBa < R_\BOb,  \\
    0,              & \text{otherwise}.
\end{cases}
\]
we infer $L_\BBa < L_\BOb$ and $R_\BBa < R_\BOb$ since no signal of binary B$_{\BB}$ is detected in the spectra. 
\[
    \mathcal{P}(L_\BBa)= 
\begin{cases}
    1,                    & \text{if } L_\BBb < L_\BBa,  \\
    0,              & \text{otherwise}.
\end{cases}
\]
\[
    \mathcal{P}(R_\BBb)= 
\begin{cases}
    1,                    & \text{if } R_\BBb < R_\BBa,  \\
    0,              & \text{otherwise}.
\end{cases}
\]
from Figure~\ref{fig:tesslc} (c), it can be observed that the eclipse of the primary depth is more than twice that of the secondary eclipse, therefore, $L_\BBb < L_\BBa$. Additionally, assuming that they are the main sequence stars, we can infer that the radius of star B$_\BBb$ is smaller than that of star B$_\BBa$.
The $\mathcal{N}$ and $\mathcal{U}$ represent the normal and uniform distributions, respectively. The posterior probability distribution for the parameters derived from the MCMC fit is shown in Figure~\ref{fig:corner_lcrv}.

The posterior distribution of the parameters $L_\BBa$, $i_\BB$, and $R_\BBb$ exhibit bimodal behavior. This bimodality arises primarily from the fact that the depth of the secondary eclipse in LC$_\BB$, which determines $L_\BBa$, is severely contaminated by errors (or de-trending) in the light curve, as shown in Figure~\ref{fig:tesslc} (c). Consequently, both $i_\BB$ and $R_\BBb$ cannot be constrained by a single model. It is also the reason why $\log g_{\BBa}$ is greater than $\log g_{\BBb}$. 

Note, we pre-calculate a grid of flux passing through the TESS filter using low-resolution spectra generated by the ATLAS9 atmosphere model\citep{Castelli2003IAUS..210P.A20C}. In our MCMC approach, we can infer $T_{\rm eff}$, $\log g$, $Z$, and $R$ for the four stars based on the parameters provided in $\theta$. Their fluxes passing through the TESS filter ($F^{\rm TESS}$) can be interpolated given $T_{\rm eff}$, $\log g$ and $Z$. Subsequently, we obtain the surface brightness through the TESS filter $L^{\rm TESS} = F^{\rm TESS} R^2$ and the surface brightness ratios of TESS filter for binary B$_\BO$ and binary B$_\BB$, 
\begin{equation}
    r_{\rm sb1} = \frac{F_{\rm TESS1b} R_\BOb^2}{F_{\rm TESS1a} R_\BOa^2}, 
\end{equation}
\begin{equation}
    r_{\rm sb2} = \frac{F_{\rm TESS2b} R_\BBb^2}{F_{\rm TESS2a} R_\BBa^2}.
\end{equation}

\begin{figure}
    \centering
    \includegraphics[width=0.95\linewidth]{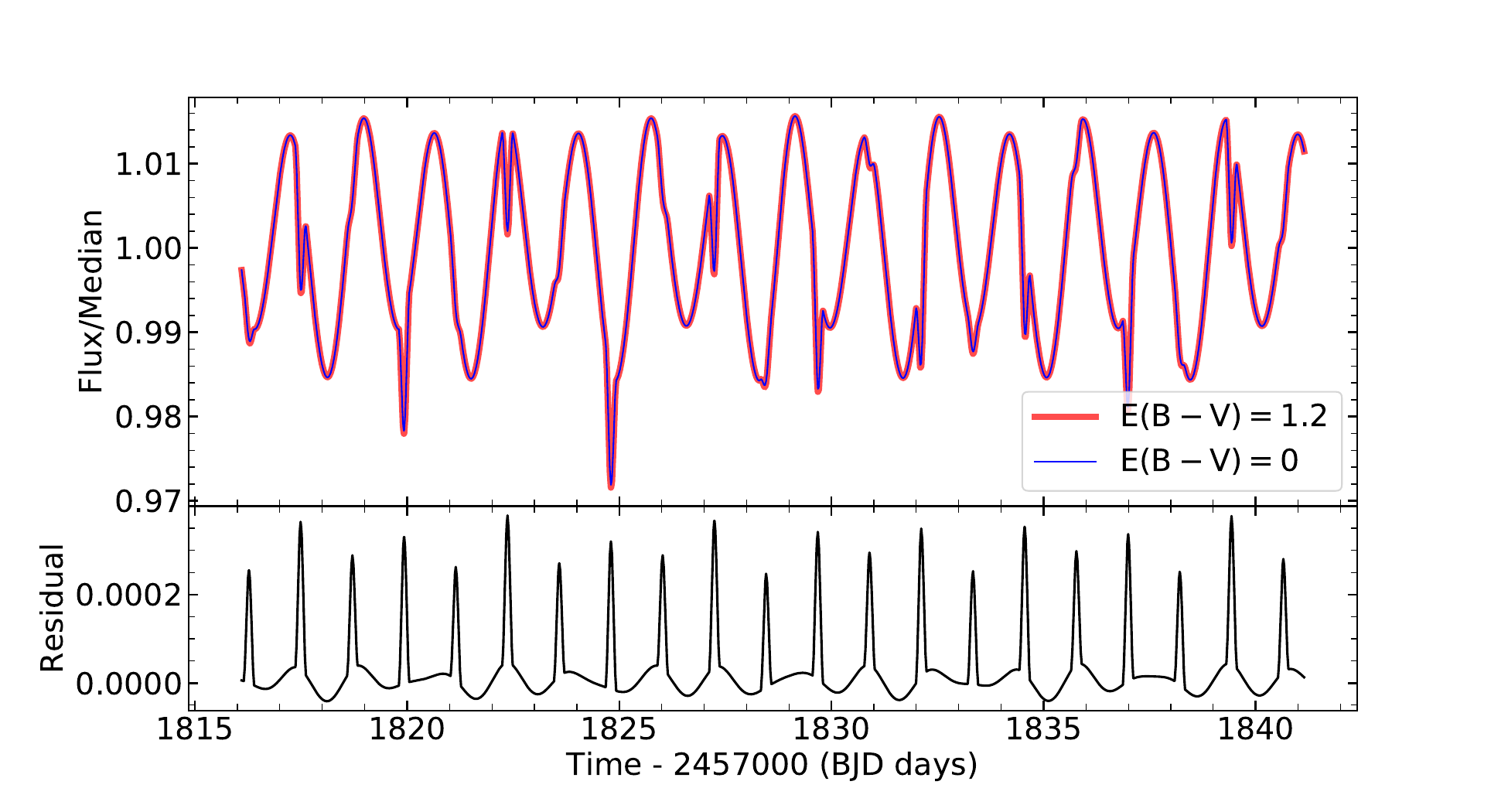}
    \caption{The model light curves of best-fit parameters but for different color excess $E(B-V) = 0$ and 1.2. The residual is calculated as the difference between the normalized fluxes $F_{E(B-V)=0} - F_{E(B-V)=1.2}$.}
    \label{fig:lcebv}
\end{figure}

The color excess $E(B-V)$ is about 1.2 estimated by the SED fitting. We calculate the surface brightness ratios at different $E(B-V) = 0$ and 1.2 and model the light curves with the best-fit parameters. We found the absolute difference in the normalized fluxes of $E(B-V)=0$ and $E(B-V)=1.2$ is less than $4\times 10^{-4}$ (see Figure~\ref{fig:lcebv}). In this case, it can be neglected that the light curve is influenced by extinction.

\begin{figure*}
    \centering
    \includegraphics[width=0.95\linewidth]{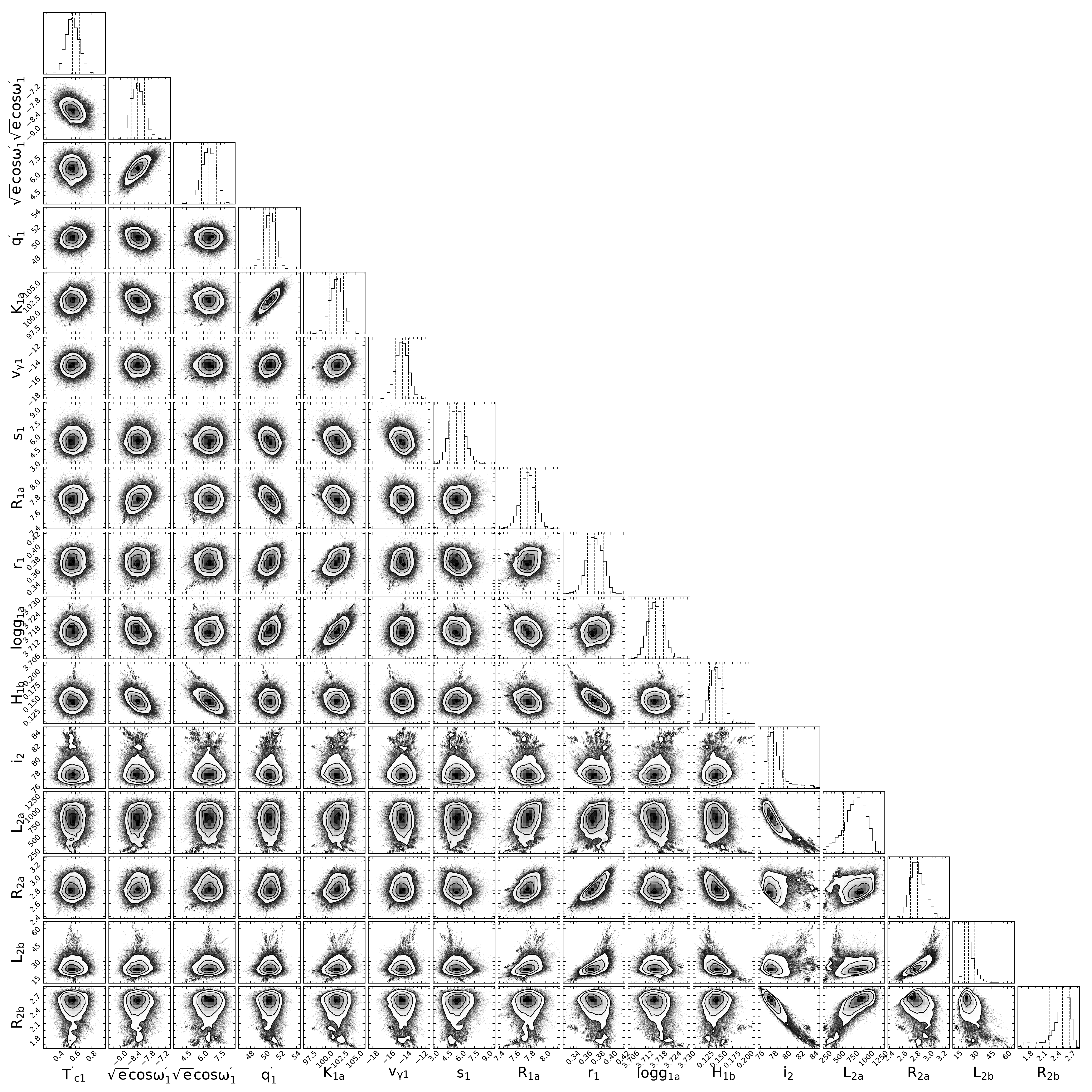}
    \caption{The posterior probability distribution for the parameters derived from the MCMC fit to the TESS light curve and radial velocity curve of binary B$_\BO$, when set the prior distribution of $\log g_\BOa$ as $\mathcal{P}(\log g_\BOa) = \mathcal{N}(3.6, 0.25)$. Note: $T'_{\rm c1} = 100(T_{\rm c1} - 2458821.51)$ is in units of BJD; $\sqrt{e}\cos \omega_{1}' = 100\sqrt{e}\cos \omega_{1}$; $\sqrt{e}\cos \omega_{1}' = 100\sqrt{e}\sin \omega_{1}$; $q'_1 = 100q_1$; $K_\BOa$, $v_{\gamma 1}$ and $s_1$ are in units of \kms; $R_\BOa$, $R_\BBa$ and $R_\BBb$ are in units of $R_\odot$; $g_{\BOa}$ in units of cm s$^{-1}$; $L_\BBa$ and $L_\BBb$ are in units of $L_\odot$; $i_2$ is in units of degree; $r_1 = R_{\BOb} R_\BOa^{-1}$; $H_{\BOb}$ is albedo effect of star B$_\BOb$.
    }
    \label{fig:corner_lcrv3.6}
\end{figure*}

The $\log g_\BOa$ value might be overestimated due to contamination of the spectrum by the other three stars (see Section~\ref{sec:spec_contaminaion}). To investigate this possibility, we conducted an experiment by setting the prior distribution of $\log g_\BOa$ as $\mathcal{P}(\log g_\BOa) = \mathcal{N}(3.6, 0.25)$ during the fitting of both the TESS light curve and the radial velocity curve. We observed that the resulting posterior distributions (see Figure~\ref{fig:corner_lcrv3.6}) closely resemble those obtained when the prior distribution of $\log g_\BOa$ is $\mathcal{N}(3.74, 0.25)$.  The derived $\log g_\BOa$ is $3.716_{-0.003}^{+0.003}$ which is consistent with the one $3.717_{-0.003}^{+0.003}$ derived from $\mathcal{P}(\log g_\BOa) = \mathcal{N}(3.74, 0.25)$.

\bibliography{quadruple}{}
\bibliographystyle{aasjournal}

\end{document}